# Detecting disruption of HER2 membrane protein organization in cell membranes with nanoscale precision

Yasaman Moradi [1,2], Jerry SH Lee [1,2,3], Andrea M. Armani [1,2]*

*armani@usc.edu

[1]University of Southern California, Mork Family Department of Chemical Engineering and Materials Science, Los Angeles, CA 90089. [2]The Lawrence J. Ellison Institute for Transformative Medicine, Los Angeles, CA 90064. [3]University of Southern California, Keck School of Medicine, Los Angeles, CA 90089

KEYWORDS: Aggregation Induced Emission, fluorescent sensor, imaging, microscopy, transmembrane proteins

**ABSTRACT:** The spatio-temporal organization of proteins within the cell membrane can affect numerous biological functions, including cell signaling, communication, and transportation. Deviations from normal spatial arrangements have been observed in various diseases, and better understanding this process is a key stepping-stone to advancing development of clinical interventions. However, given the nanometer length scales involved, detecting these subtle changes has primarily relied on complex super resolution and single molecule imaging methods. In this work, we demonstrate an alternative fluorescent imaging strategy for detecting protein organization based on a material that exhibits a unique photophysical behavior known as aggregation induced emission (AIE). Organic AIE molecules have an increase in emission signal when they are in close proximity and the molecular motion is restricted. This property simultaneously addresses the high background noise and low detection signal that limit conventional widefield fluorescent imaging. To demonstrate the potential of this approach, the fluorescent molecule sensor is conjugated to a human epidermal growth factor receptor 2 (HER2) specific antibody and used to investigate the spatio-temporal behavior of HER2 clustering in the membrane of HER2-overexpressing breast cancer cells. Notably, the disruption of HER2 clusters in response to an FDA-approved monoclonal antibody therapeutic (Trastuzumab) is successfully detected using a simple widefield fluorescent microscope. While the sensor demonstrated here is optimized for sensing HER2 clustering, it is an easily adaptable platform. Moreover, given the compatibility with widefield imaging, the system has the potential to be used with high-throughput imaging techniques, accelerating investigations into membrane protein spatio-temporal organization.

Transmembrane proteins play a critical role in governing fundamental cell processes like cell signaling and cell division. However, in many cases, it is not simply the presence or absence of a given protein but their spatio-temporal organization within the membrane that modulates biological processes[1,2]. Therefore, the ability to decipher these dynamic interactions is key to unravelling how they mediate the cellular signal transductions which play a role in a range of health conditions.

At a fundamental science level, several methods exist for measuring the spatial organization of membrane receptors including electron microscopy[3,4], optical and fluorescence microscopy-based techniques[5–10], and proximity-based assays[11–13]. Among these, electron microscopy provides the highest resolution. However, because it is not compatible with live cells, it can only provide static snapshots of dynamic processes which do not fully capture the behavior of receptors in the membrane. Alternatively, super resolution imaging technologies can be used. However, these are not conducive to high throughput sample analysis methods and rely on extremely specialized instrumentation[14]. Given the complexity of these interactions, the ability to acquire large data sets and perform high-dimensional analysis is key to unraveling the underlying biological control mechanisms.

To overcome these challenges requires re-thinking our approach to imaging. One strategy is to develop new types of fluorescent probes which can serve as proximity sensors and operate in a high throughput system, and one promising material system is based on organic molecules that exhibit Aggregation Induced Emission (AIE) behavior.

Aggregation Induced Emission (AIE) describes a photophysical property in which the fluorescent intensity of a fluorophore aggregate is higher than when the molecule is well-dispersed in solution[15]. This behavior is attributed to a restriction of the intramolecular rotations which increases the emissivity of the molecule[16,17]. In contrast, most commercially available fluorescent compounds exhibit the opposite behavior due to strong $\pi - \pi$ interactions, and their emission is reduced or completely quenched when they aggregate in solid state or are in high concentration solutions[17,18]. To date, AIE molecules have been used for biomolecule sensing in solution[19–22], bioimaging[23–28], monitoring protein folding/ unfolding processes in cells[29], and sensing the interaction of proteins in solution[30–32]. Therefore, a material exhibiting an AIE response has the potential of being utilized for studying spatial arrangement of membrane receptors at the nanometer length scale using conventional fluorescent microscopy.

In this work, we develop a fluorophore based on tetraphenylethylene (TPE), an AIE molecule, for studying the clustering behavior of HER2 in HER2-overepressing breast cancer cells and investigate the response of HER2 cluster to a therapeutic. HER2 is involved in the regulation of several signaling pathways that lead to an increase in cell proliferation in several types of cancer[33,34]. Because HER2 is known to localize in clusters on the membrane[3,35,36], one therapeutic strategy is the disruption of this organization[37,38]. To study this dynamic nanoscale process, the TPE-based probe is conjugated to an antibody specific to the extracellular domain of HER2. As shown in **Figure 1a**, when the HER2 proteins are separated, the developed fluorophore sensor is not emissive. However, when the concentration of HER2 increases and clusters form, the molecule undergoes a fluorescent turn-on process due to the AIE behavior of the TPE (**Figure 1b**). Because only HER2 molecules in close proximity initiate the turn-on process, this approach overcomes many of the previous limitations by providing a method to detect HER2 clustering and simultaneously reducing background noise. Furthermore, since AIE is a reversible process, the dynamic interactions of HER2 proteins in response to external stimuli, like therapeutics, can be studied.

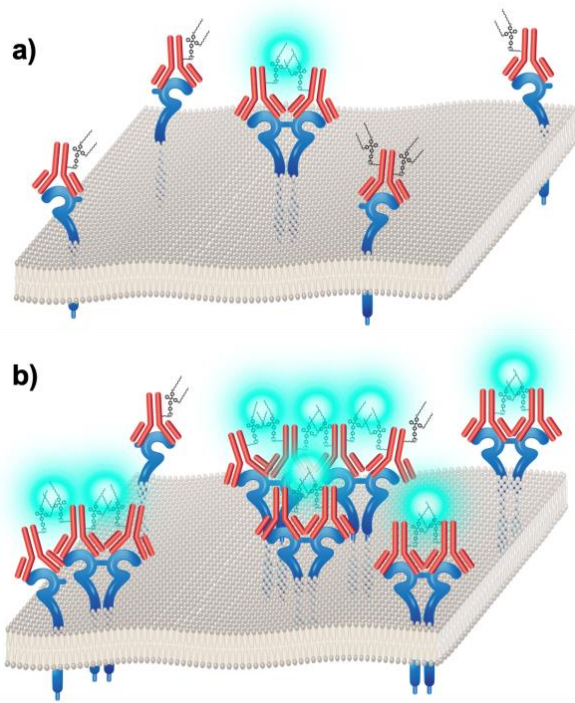

**Figure 1.** Schematic of AIE sensor for localized detection and visualization of HER2 clusters or HER2s that are localized in each other's proximity on **(a)** HER2-negative (low HER2 expression) and **(b)** HER2-positive (overexpressing) cell surfaces.

## EXPERIMENTAL SECTION

The detailed information of the chemicals and reagents, instrumentation, synthesis, bioconjugation, and characterization results of the intermediates and final molecule as well as the validation of the cell lines are presented in the supplementary information.

**Synthesis and characterization of AIE sensor.** The synthetic details and NMR spectra confirming the synthesis of TPE-NHS are included in the SI. Post synthesis, the absorption and emission spectrum of TPE-NHS were characterized in distilled water, DMSO, and DMEM cell media. Furthermore, the AIE response of TPE-NHS molecule was confirmed. Lastly, the potential for Trastuzumab to directly interact with TPE-NHS was studied. All details and results are included in SI.

To develop the AIE sensor, the TPE-NHS molecule was conjugated to HER2 specific antibody through a micelle mediated bioconjugation reaction that leveraged the intrinsic Lysine residues of the HER2 specific antibody. The details of the bioconjugation reaction, purification of free TPE-NHS using gel spin column, conjugation confirmation, and optical characterization methods are included in SI.

**Direct immunofluorescent imaging.** SKBR3 cells (ATCC, HTB-30) and MCF7 cells (ATCC, HTB-22) were seeded at the density of 7,000 cells per well in a 96 well glass-bottom plate (Cellvis, P96-0-N) and incubated for 3 days before running the assay to reach the approximate confluency of 70%. After 3 days, the media was removed, and the cells were fixed using 4% Paraformaldehyde (Alfa Aesar, J62478) and blocked using 2% BSA blocking buffer (Thermo scientific 37525) for one hour. After washing the fixed cells 3 times (5 minutes each), the antibody dye conjugates (Fluorescein-HER2 Ab as the positive control and different TPE-HER2 Ab conjugates) were diluted to the concentration of 10 μg/ml in 0.1% BSA solution, added to the fixed SKBR3 and MCF7 cells, and left at 4°C overnight. The samples were washed with 1x PBS 2 times and imaged on Zeiss Axio observer connected to a X-cite Series 120Q light source using the excitation and emission filters of interest using a 20x objective. The TPE filter cube has the excitation of G365, BS of 395, and emission BP of 535/30. The Fluorescein filter cube has the excitation BP of 500/25, BS of 515, and emission BP of 535/30.

**Colocalization immunofluorescent imaging.** SKBR3 cells were seeded at the density of 7,000 cells per well in a 96 well glass-bottom plate (Cellvis, P96-0-N) and incubated for 3 days before running the assay to reach the approximate confluency of 70%. After 3 days, the media was removed, and the cells were fixed using 4% Paraformaldehyde (Alfa Aesar, J62478) and blocked using 2% BSA blocking buffer (Thermo scientific 37525) for one hour. After washing the fixed cells 3 times, staining solution consisting of 20 μg/ml of TPE-HER2 Ab and 20 μg/ml Texas red-HER2 Ab (Invitrogen, T20175) in 0.1% BSA was added to the cells. After overnight staining at 4°C, fixed SKBR3 cells were washed and imaged in the brightfield channel, TPE channel (filter cube with the excitation of G365, BS of 395, and emission BP of 535/30) and Texas red channel (filter cube with the excitation BP of 550/25, BS of 570, and emission BP of 605/70) on Zeiss Axio-observer widefield fluorescent microscope using a 20x objective.

**Analysis of the impact of Trastuzumab on HER2 clusters.** SKBR3 cells (ATCC, HTB-30) were seeded at the density of 7,000 cells per well in a 96 well glass-bottom plate (Cellvis, P96-0-N) and incubated for 3 days. After reaching the ideal cell confluency on day 3, different concentrations of Trastuzumab (Selleckchem, A2007) in McCoy media (from 0 μg/ml to 100 μg/ml) were prepared by serial dilution. The old cell media was replaced with the Trastuzumab media and incubated for 2, 8, and 24 hours. Then, cells were fixed and washed following the direct immunofluorescent protocol. TPE-HER2 Ab and Fluorescein-HER2 Ab at the concentrations of 10 μg/ml in 0.1% BSA solution was added to the fixed cells and left at 4°C overnight. The samples were washed with 1x PBS 2 times and imaged on Zeiss Axio observer using a 20x objective. The TPE filter cube with the excitation of G365, BS of 395, and emission BP of 535/30 and the Fluorescein filter cube with the excitation BP of 500/25, BS of 515, and emission BP of 535/30 was used. The imaging parameters were held constant for all of the wells, allowing this value to be used for comparison across conditions. Post-imaging the data was analysed based on the description in the Image analysis method section.

**Image analysis.** An in-house developed image analysis tool which is available on GitHub was used to track the fluctuations in the fluorescent signal post treatment. The image analysis tool uses an edge detection algorithm to define the region of interest (ROI) which was the cell area. Then, the defined mask was applied to the fluorescent image. Summation of the pixel values of the masked region and the total masked area (number of the pixels) was extracted from each fluorescent image. Dividing the summation of the pixel values by the masked area, a value representing the average fluorescent intensity per fluorescent image was calculated. The detailed description of the image analysis platform is included in SI.

## RESULTS AND DISCUSSION

**Design rationale of the AIE molecule.** The primary design criteria for the fluorescent sensor are maintaining high biological specificity while achieving nanoscale spatial sensitivity. One common strategy to endow specificity to a fluorescent probe is to conjugate the molecule directly to the amine groups that are located on the Fc-region of the antibody[39]. In this work, the TPE-based molecule was functionalized with N-Hydroxy succinimide (NHS) ester which subsequently bound to a monoclonal antibody specific to HER2[39,40]. In addition, a pair of hydrocarbon chains were added to increase the probability of intermolecular restriction as the HER2 proteins cluster.

The length of the TPE-NHS was modelled using density functional theory (QChem). This value sets the minimum and maximum interaction distance between a pair of TPE molecules that could give rise to the AIE response. Based on the results, detection of TPE-NHS is in the range of 2 nm- 6 nm (**Figure 2**). Additional details on the modelling are presented in the SI (**Figure S1**).

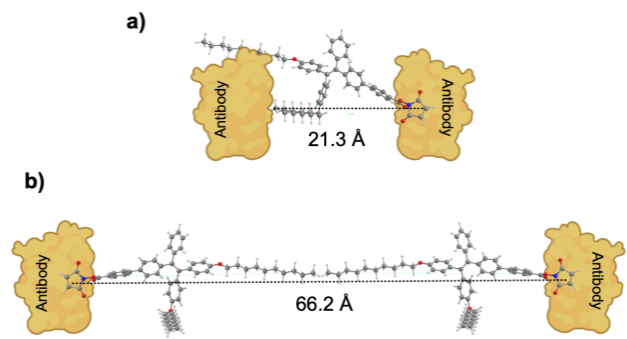

**Figure 2.** Distance of the NHS ester portion of TPE-NHS from the end of the hydrocarbon chains in the ground state can determine the detection range of the AIE sensor. The TPE- bound antibodies binding to HER2 protein can detect **(a)** a minimum distance of 21.3 Å upon interaction of the TPE-NHS molecule from the shorter side with the surface of a nearby HER2 bound antibody and **(b)** the maximum distance of 66.2 Å upon interaction of two molecules from their longest length.

**Synthesis and characterization of the AIE molecule.** The synthesis process of TPE-NHS molecule is shown in Scheme 1a[41–43], and the different chemical groups of TPE-NHS are color-coded to highlight their role in the molecule's operation. The intermediate and final structures were confirmed with NMR (**Figure S2-S9**).

The spectral properties (absorption and emission) of TPE-NHS were characterized in a range of solvents with different polarities. According to the results in **Figure S10**, TPE-NHS demonstrates slight solvatochromic shifts in the absorbance and more significant shifts in the emission spectrum in DMSO, distilled water, and DMEM cell media. Previous work has shown that solvatochromism in organic molecules can be attributed to the stabilization of the electronic excited state of the molecule by the polar solvent, and this response is largely attributed to the hydrogen bonding abilities of the specific solvent[44]. However, other solvent properties can play a role. The observed results align with prior studies of solvatochromism in AIE molecules[45–47].

Solubility testing confirmed that the amphiphilic TPE-NHS is soluble in mildly polar Dimethyl sulfoxide (DMSO) which has a relative polarity of 0.44[48]. The AIE response of TPE-NHS was confirmed using two approaches. First, increasing the concentration of the molecule in the 99 (v/v) % solution of distilled water/DMSO results in the initiation of aggregation induced fluorescence at concentrations above 10 μM (**Figure S11a**). Second, by increasing the relative volume ratio of distilled water:DMSO, the TPE-NHS begins aggregating, and the fluorescent emission in the system increases (**Figure S11b**).

The proposed application of this sensor is detecting the effect of Trastuzumab on HER2 clustering using the AIE response of TPE. Therefore, it is important to ensure that Trastuzumab does not directly interact with the fluorescent behavior of the TPE. A series of experiments were performed using the same range of Trastuzumab concentrations as in the cell line studies. No effect on the AIE response was observed (**Figure S12**). Additional details are in the SI.

**AIE sensor development and characterization.** To endow specificity to the TPE-NHS dye, it was conjugated to an antibody specific to the extracellular domain of the HER2. This conjugation process relied on the NHS ester bonding with the amines that are part of the lysine residue, forming an amide bond with the HER2 antibody. Due to the potential of antibodies to denature and loose biological specificity, the NHS ester-Amine conjugation chemistry is performed in aqueous solutions (**Scheme 1b, Figure S13**). To reduce the formation of aggregates and increase the bioconjugation yield, a micelle-mediated conjugation process using Tween-20 was performed[49]. To optimize the Tween-20 concentration in the reaction solution, a series of TPE-NHS/HER2 antibody conjugation reactions using a range of Tween-20 concentrations were performed. Based on the results from the optimization study (**Figure S14**), 0.025 (v/v) % of Tween-20 was used in the antibody bioconjugation process.

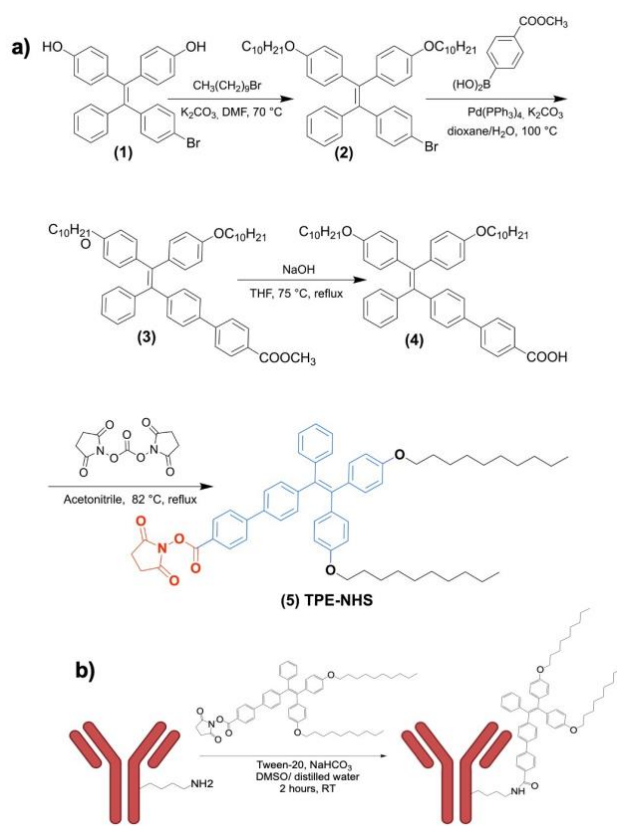

**Scheme 1.** **(a)** Schematic of synthesis of TPE-NHS (compound 5). The three key components of TPE-NHS are indicated in red (NHS Ester), blue (AIE group), and black (alkane chain) and **(b)** Schematic of the AIE sensor development by conjugation of TPE-NHS to the lysin residue of antibody.

To confirm the conjugation of TPE-NHS with the HER2 antibody, MALDI mass spectroscopy was used. Results presented in **Figure S15** show a simultaneous shift and broadening of the peak after the antibody conjugation process is performed. The software reported a 1.015 kDa shift in the m/z values of HER2 antibody and TPE-HER2 Ab (TPE-NHS conjugated to HER2 Antibody) spectrum. Considering the fact that the expected molecular weight of the each TPE-NHS post conjugation is 749 g/mol, this result demonstrates the conjugation of TPE to the HER2 antibody with an estimated average dye to antibody ratio of 1.35. Furthermore, a complementary SDS-PAGE assay further confirmed the formation of the TPE-HER2 Ab **(Figure S16)**.

After conjugation, the optical absorption and emission wavelengths of the developed AIE sensor were characterized in 1x PBS. The absorption maximum of the conjugated molecule is 354 nm, and the emission wavelength is centered at 518 nm (**Figure S17**) which indicates a clear Stokes shift of 164 nm. Importantly, these results confirm that the fluorescent behavior of the dye has not been significantly altered by the bioconjugation process.

**Evaluation of the AIE sensitivity, cytotoxicity, and HER2 target specificity of the AIE sensor (TPE-HER2 Ab).** The aggregation induced emission behavior of TPE-HER2 Ab was analysed by measuring the emission over a range of sample concentrations in 1x PBS **(Figure 3a, Figure S18)**. Results in **Figure 3b** show an increasing trend in the fluorescent emission intensity at the concentrations above 0.04 mg/ml which confirms that expected AIE behavior of the AIE probe is maintained after antibody conjugation.

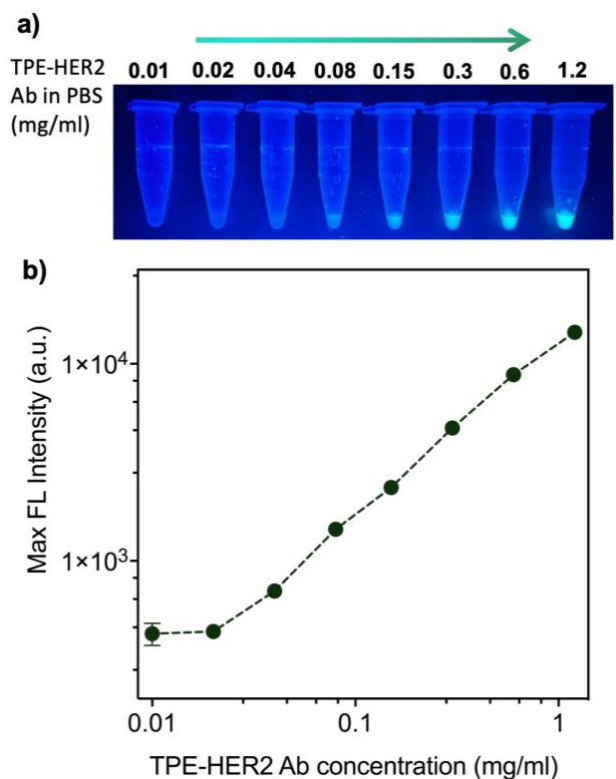

**Figure 3.** Fluorescent behavior of TPE-HER2 Ab. **(a)** Optical images of different concentrations of TPE-HER2 Ab in 1x PBS excited by a 365 nm UV lamp. **(b)** Maximum FL intensity at each concentration of TPE-HER2 Ab in 1x PBS plotted against the concentration of TPE-HER2 Ab in each sample. In some cases, the error bars are not visible because they are smaller than the symbols.

Before applying TPE-HER2 Ab in cell imaging applications, cytotoxicity of TPE-NHS probe was evaluated on two different breast cancer cell lines, SKBR3 and MCF7, using concentrations up to 50 μM. SKBR3 over-expresses HER2, and MCF7 is a HER2- low expressing cell line[50]. The HER2 expression levels of these cell lines were confirmed using indirect immunofluorescent staining and western blot (**Figure S19**). Based on the Cell Titer Glo (CTG) assay results, no significant impact on cell viability was observed over the entire range of TPE-NHS concentrations studied as compared to the positive and negative controls (**Figure S20**). This observation confirms the low cytotoxicity of the compound and its potential applicability in live cell imaging studies.

In order for the TPE-HER2 Ab to accurately monitor the HER2 clustering process, it must selectively bind to HER2. The specificity of the TPE-HER2 Ab is evaluated by performing a colocalization, competition fluorescent imaging measurement using the SKBR3 cell line. Texas red was selected as the second fluorophore because the excitation (emission) wavelengths of Texas red do not overlap with the excitation (emission) wavelengths of TPE, allowing for colocalization to be easily determined by merging the fluorescent images (**Figure S22**). Both fluorophores were conjugated to the same type of HER2 antibody, removing the variability of antibody binding site and affinity from the measurement. An identical series of colocalization imaging measurements are performed using Fluorescein in place of the TPE.

Brightfield and fluorescent images of the HER2-overexpressing SKBR3 cells labelled with Texas red-HER2 Ab and TPE-HER2 Ab are shown in **Figure 4a-c**. The emission signal from each fluorophore is easily observable. One approach to quantify the signal quality is the signal-to-noise ratio (SNR). According to basic signal processing theory, an SNR above 1 is considered detectable. For the present images, the SNR of TPE and Texas-red are 10.23± 0.50 and 20.66 ± 2.22, respectively. Therefore, both fluorophores are providing robust detection signals. The details of SNR calculation are included in the SI.

When the Texas red and TPE fluorescent images are merged (**Figure 4d**), the colocalization of the TPE and the Texas red is qualitatively evident. The same colocalization experiment using Fluorescein-HER2 Ab and Texas-red HER2 Ab was also performed as a control measurement, and similar results were obtained (**Figure S23**).

To quantitatively analyse colocalization, the photon intensities in the Texas red and the TPE channels are spatially correlated (**Figure 4e**). This analysis is also performed for the previously discussed Fluorescein control measurement (**Figure 4f**). Furthermore, the Pearson Correlation Coefficient (PCC) of the TPE:Texas red HER2 Ab and the Fluorescein: Texas red HER2 Ab were calculated[51]. Across multiple images, the PCC values of the TPE:Texas red and the Fluorescein:Texas red channels were calculated to be 0.74± 0.06 and 0.65± 0.03, respectively (**Figure S24**). Therefore, the TPE-HER2 Ab was able to specifically target HER2 on SKBR3 cells, setting the stage for HER2 cluster detection and analysis using this AIE-based imaging probe.

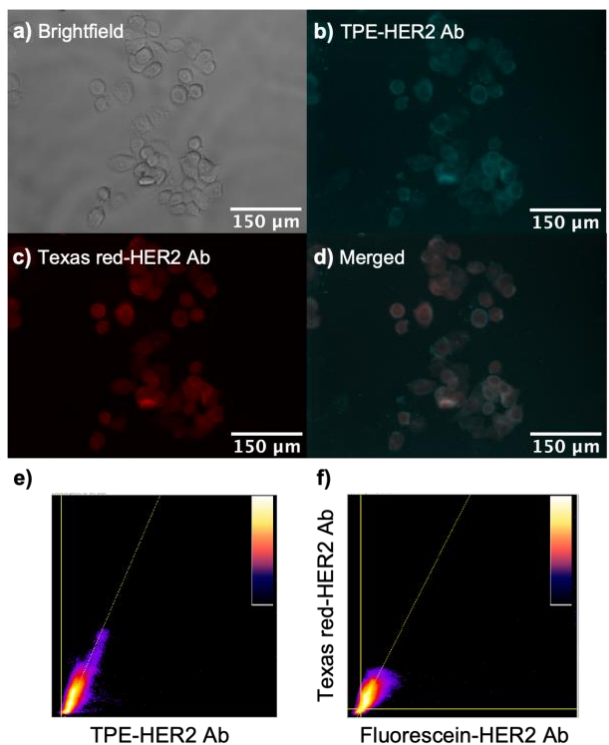

**Figure 4.** Colocalization competition analysis of TPE-HER2 Ab and of Texas red-HER2 Ab in SKBR3 (HER2-overexpressing) cells. Images of SKBR3 cells stained with 20 μg/ml of TPE-HER2 Ab and 20 μg/ml of Texas red-HER2 Ab are shown in **(a)** brightfield channel, **(b)** TPE fluorescent channel, **(c)** Texas-red fluorescent channel, and **(d)** merged channel of TPE and Texas red fluorescent channels. 2D intensity histogram of **(e)** TPE: Texas red channels and **(f)** Fluorescein: Texas red channels for photon intensity correlation analysis.

**AIE sensor (TPE-HER2 Ab) for detection of HER2 clustering in cancer cells.** The TPE-HER2 Ab imaging agent was used to detect HER2-HER2 interactions and HER2 clustering in two different breast cancer cell lines (SKBR3 and MCF7).

One challenge with antibody-based assays is batch-to-batch variations in antibody reactivity. To remove this variable and potential measurement confound, all measurements were performed with the same antibody lot[52]. As shown in **Figure 5a-c**, the fluorescent emission signal from the TPE in the HER2-overexpressing SKBR3 cells was clearly identifiable with an SNR of 9.67± 1.48. This signal is directly related to the AIE fluorescent mechanism of the TPE moiety. Namely, as the HER2 clusters form, the TPE molecular motion becomes restricted, and the fluorescent intensity increases. In contrast, in the HER2-low expressing MCF7 cells, the fluorescent signal is barely detectable with an SNR of 6.13± 0.76. (**Figure 5d-f**). This difference is due to a lack of HER2 clustering as well as low HER2 concentration, and it is in agreement with the cell line validation assays.

Taken together, these observations in the SKBR3 and MCF7 cell lines confirm the ability of the TPE-HER2 Ab imaging agent to selectively target HER2 in cells and to detect HER2 clusters in the cell membrane. These findings set the stage for monitoring the dynamic process of HER2 cluster formation and disruption.

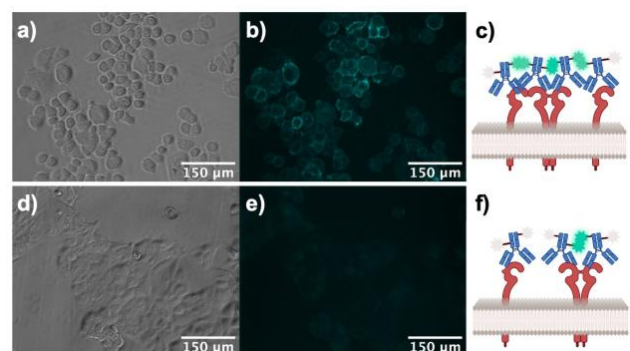

**Figure 5.** Brightfield and fluorescent images of **(a, b)** SKBR3 (HER2 overexpressing) and **(d, e)** MCF7 (HER2- low expressing) cells stained with 10 μg/ml of TPE-HER2 Ab along with the illustration of the TPE mediated fluorescent turn on process on the surface of **(c)** SKBR3 and **(f)** MCF7 cells.

**AIE sensor for analyzing the impact of cancer cell therapeutic treatment on HER2 clusters.** Trastuzumab (Herceptin) is a humanized monoclonal antibody that has been approved by the Food and Drug Administration (FDA) for patients with HER2-positive invasive breast cancer[53]. This therapeutic is a HER2 specific antibody that binds the juxtamembrane portion of the HER2 extracellular domain[54,55]. Several mechanisms of action have been proposed including prevention of HER2-receptor dimerization[54,56,57]. However, it is not clear if dimerization inhibition alone is sufficient to achieve the therapeutic results that are observed[54,58]. Another hypothesis is that changing the biophysical pattern and distribution of HER2 clusters on the cell membrane gives rise to the observed effect[55,59]. The TPE-HER2 Ab imaging agent developed here is uniquely suited to provide insight into this scientific question.

SKBR3 cells were seeded in a 96 well plate in triplicate and were treated with Trastuzumab concentrations ranging from 0 μg/ml to 100 μg/ml for 2, 8, and 24 hours. The concentration and time ranges were based on prior Trastuzumab studies[59–61]. Subsequently, the TPE-HER2 Ab imaging agent was added to the plates following the developed direct immunofluorescent staining protocol. In parallel, Fluorescein-HER2 Ab was used as a control. Fluorescent imaging was performed using a widefield fluorescent microscope with a 20x objective, and the images were analysed and quantified using an in-house developed computational method. Details are included in the SI (**Figure S25**).

In the assay using the Fluorescein-HER2 Ab, the average fluorescent intensity does not noticeably change in any of the Trastuzumab incubation times or concentrations used (**Figure 6a-c, Figure S26**). Specifically, the SNR varies from 15.90± 0.90 to 15.74± 1.80 over the course of the 24-hours imaging measurement, which is within the error of the SNR values. Therefore, even with a robust SNR value, there is no detectable change. Because Fluorescein-HER2 Ab is not sensitive to the proximity of the receptors, this result indicates that the HER2 expression is constant and the absolute concentration of HER2 in the cell membrane is not changed. This result is expected. However, it does not provide insight into the mechanism of action of the Trastuzumab and its impact on the HER2 clustering process.

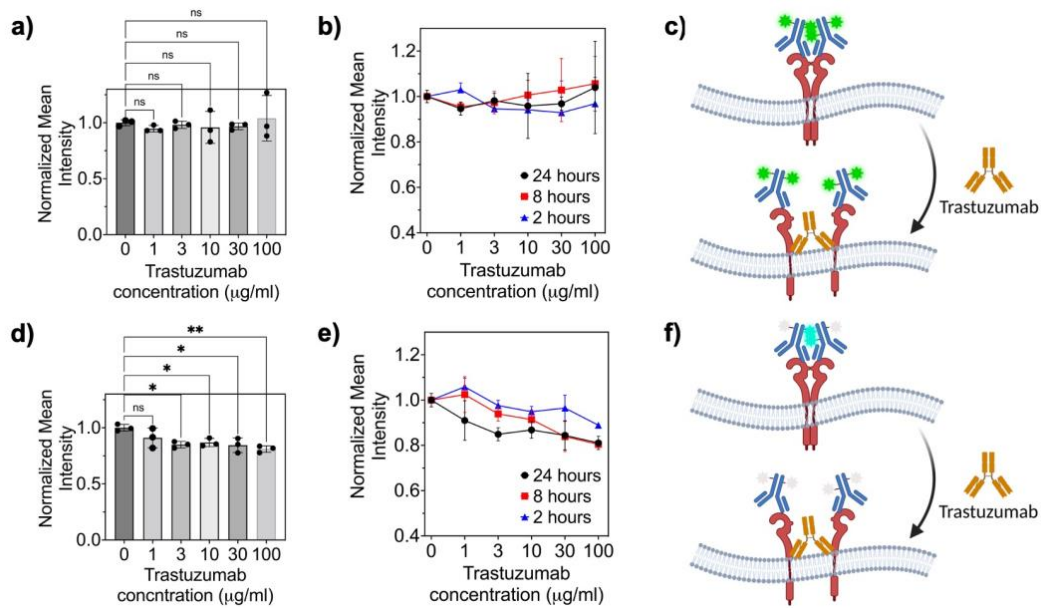

**Figure 6.** Normalized mean fluorescent intensity of SKBR3 cells stained with **(a, b)** 10 μg/ml Fluorescein-HER2 Ab and **(d, e)** 10 μg/ml TPE-HER2 Ab after treatment with a range of concentrations of Trastuzumab (0 μg/ml to 100 μg/ml) for **(a, d)** 24 hours and **(b, e)** different time intervals of 2 hours, 8 hours, and 24 hours all in one plot. The data is collected in triplicate, and on average, each data point consists of 30 SKBR3 cells. (* $p < 0.05$ and ** $p < 0.01$). Schematic of HER2 overexpressing SKBR3 cell membrane indicating the fluorescent response of **(c)** Fluorescein-HER2 Ab and **(f)** TPE-HER2 Ab stained HER2 proteins after Trastuzumab treatment.

In contrast, by monitoring the fluorescence intensity, the TPE-HER2 Ab provides information about the HER2 clustering behavior. With 2 and 8 hours of Trastuzumab treatment, a decrease in cluster formation is observed only at highest concentrations (**Figure 6e, Figure S26**). However, with 24 hours of treatment, the clustering process is reduced with all concentrations above 1μg/ml studied in this work (**Figure 6d-e**). In comparing the SNR values over the course of the 24-hours imaging measurement, the SNR changes from 11.85± 0.55 to 10.73± 0.57.

This response to Trastuzumab is in agreement with prior studies exploring the impact of Trastuzumab on HER2 clusters in HER2 overexpressing cells[58]. This result reveals that the decreasing fluorescent trend is due to the interaction between the Trastuzumab and the HER2 clusters. These findings confirm the capability of our AIE sensor (TPE-HER2 Ab) for the detection and visualization of HER2 clustering dynamics upon exposure to an external stimulus like therapeutics.

## CONCLUSIONS

In summary, we developed a targeted imaging agent based on the combination of an AIE fluorophore with a monoclonal antibody and used it to monitor the clustering process of HER2 in response to a HER2-targeting therapeutic (Trastuzumab). The TPE- HER2 Ab conjugate system is specific to the target and can visualize the HER2 clustering process in HER2 overexpressing cells. Furthermore, using the developed fluorescent probe, the impact of Trastuzumab on the proximity of HER2 in the SKBR3 cells can be readily visualized and determined. The results demonstrate an easily generalizable method for studying the distribution and nanoscale localization of membrane-bound proteins. Given the target specificty of TPE-HER2 Ab and biocompatibility of TPE-NHS, this AIE sensor has the potential of being optimized for studying the dynamics of HER2 clusterization in live cells. Furthermore, while the focus of this work is on using HER2 antibodies, by shifting to peptide or similar engineered targeting moieties, quantification of the signal could be possible. This additional capability could lead to applications in developing improved therapeutics or in understanding membrane dynamics[62,63].

## ASSOCIATED CONTENT

### Supporting Information

DFT modeling of TPE-NHS, detailed synthesis procedures and characterization, UV-vis absorption and fluorescence spectra of TPE-NHS, Effect of Trastuzumab on TPE-NHS emission, bioconjugation optimization imaging, MALDI and SDS-PAGE bioconjugation characterization, UV-vis absorption and fluorescence spectra of TPE-HER2 Ab, Cell culture procedures, Cell line validation through immunofluorescent imaging and western blot, Cell viability assay, SNR calculation protocol description, Control multichannel fluorescent imaging, Control colocalization assay, Pearson Correlation Coefficient data, detailed image analysis protocol description, data for the AIE based trastuzumab treatment assay.

## AUTHOR INFORMATION

### Corresponding Author

* armani@usc.edu

### Author Contributions

Andrea M. Armani and Yasaman Moradi were responsible for experimental design and data analysis. Yasaman Moradi performed the experiments. Andrea M. Armani, Jerry SH Lee, and Yasaman Moradi wrote and revised the manuscript. Andrea M. Armani and Jerry SH Lee supervised the project. All co-authors have given approval to the final version of the manuscript.

### Funding Sources

This work was supported by the Office of Naval Research (N00014-22-1-2466, N00014-21-1-2044), Uniformed Services University of the Health Sciences (USUHS) award from the Defense Health Program to the Murtha Cancer Center Research Program administered by the Henry M. Jackson Foundation for the Advancement of Military Medicine (HU00011820032), and the Ellison Institute of Transformative Medicine. Disclaimer: The contents of this publication are the sole responsibility of the authors and do not necessarily reflect the views, opinions, or policies of the USUHS, the Henry M. Jackson Foundation for Advancement of Military Medicine, Inc., the Department of Defense, the Department of the Army, Navy, or Air Force. Mention of trade names, commercial products, or organization does not imply endorsement by the U.S. Government.

**Notes**

J.S.H.L. serves as Chief Science and Innovation Officer for Ellison Institute, LLC (paid); board of trustee for Health and Environmental Institute, Inc. (unpaid, travel support); and scientific advisory board for AtlasXomics, Inc., and ATOM, Inc. (unpaid, travel support). A.M.A. serves as the Senior Director of Engineering and Physical Science for Ellison Institute, LLC (paid).

ACKNOWLEDGMENT

The authors would like to thank Soheil Soltani for development of the image analysis MATLAB code, Yingmu Zhang for assisting with chemical design, Jonathan Katz for assisting with Mass spectroscopy experimental design, the Multi-omics Mass Spectrometry Core (MMSC) at USC School of Pharmacy for MALDI data acquisition, and Kian Kani and Carolina Garri for assisting with bioassay designs.

ABBREVIATIONS

AIE, Aggregation Induced Emission; TPE, tetraphenylethelene; HER2, human epidermal growth factor receptor 2; DMSO, dimethyl sulfoxide; DMEM, Dulbecco's modified eagle medium; BSA, bovine serum albumin; Ab, antibody; PBS, phosphate-buffered saline; BP, band-pass, NMR, nuclear magnetic resonance; MALDI, matrix assisted laser desorption/ionization; SDS-PAGE, Sodium dodecyl-sulfate polyacrylamide gel electrophoresis; CTG, Cell Titer Glo; SNR, Signal-to-Noise Ratio; PCC, Pearson Correlation Coefficient; FDA, Food and Drug Administration.

# Detecting disruption of HER2 membrane protein organization in cell membranes with nanoscale precision

Yasaman Moradi[1,2], Jerry SH Lee[1,2,3], Andrea M. Armani[1,2*]

[1]University of Southern California, Mork Family Department of Chemical Engineering and Materials Science, Los Angeles, CA 90089. [2]The Lawrence J. Ellison Institute for Transformative Medicine, Los Angeles, CA 90064. [3]University of Southern California, Keck School of Medicine, Los Angeles, CA 90089
*armani@usc.edu

Table of Contents

## 1   Density functional theory (DFT) modeling of TPE-NHS

The ground state equilibrium geometry of TPE-NHS was calculated by density functional theory (DFT) in the gas phase at the B3LYP/6-311g* level of theory. Then, the length of the molecule from the NHS ester part to the end of the each of hydrocarbon chains were measured in Chem3D. Results in **Figure S1a** and **Figure S1b** demonstrates these lengths to be 33.1 Å and 21.3 Å. These values can also estimate the HER2 proximity detection of the AIE molecule to be in the range of 2.2 nm- 6.6 nm.

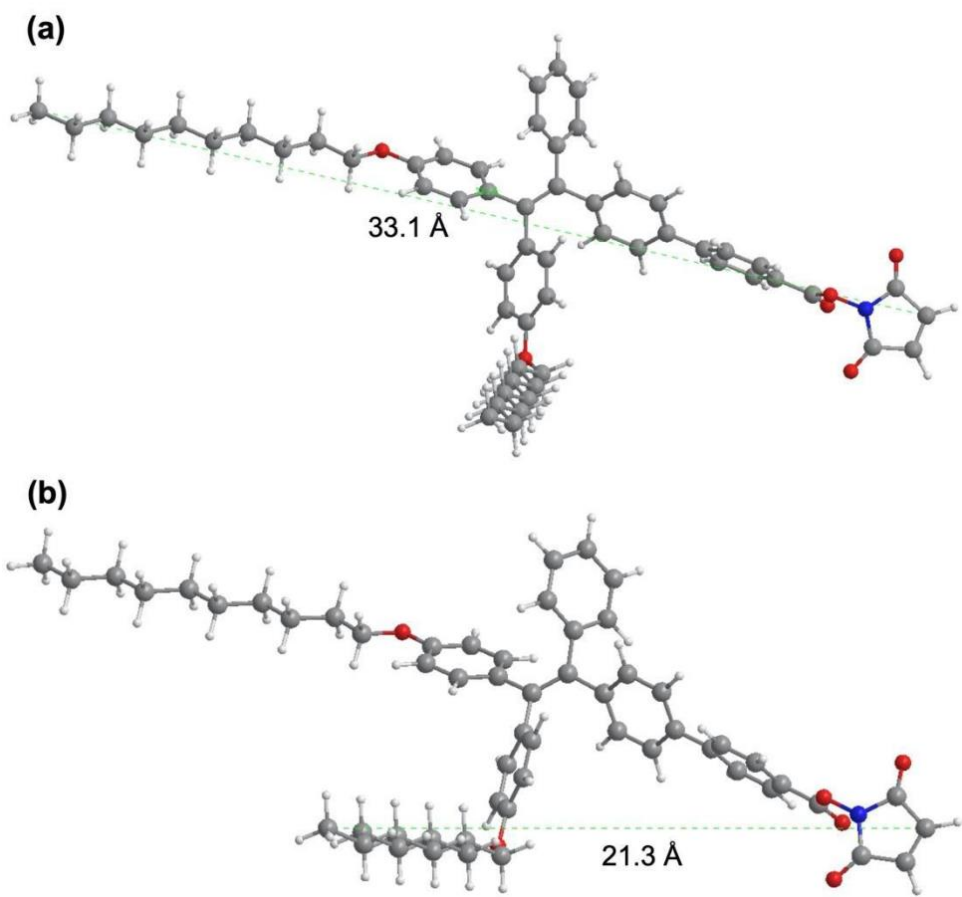

**Fig S1. (a)** Maximum and **(b)** minimum distance of the NHS ester portion of TPE-NHS from the end of the hydrocarbon chains in the ground state.

## 2   Synthesis of TPE-NHS

TPE-NHS was synthesized following the reaction path in the manuscript. The detail of each reaction is explained in each section followed by the $^1$H NMR and $^{13}$C NMR confirmation data.

All chemical reagents and solvents were either purchased from VWR or Sigma Aldrich. $^1$H NMR and $^{13}$C NMR spectra were recorded on a Varian Mercury 400 MHz spectrometer with 96-spinner sampler changer using either deuterated chloroform or DMSO as solvent, as indicated.

**Synthesis of 1.** Compound **1** was synthesized and purified based on a previously published protocol[1].

**Synthesis of 4,4'-(2-(4-bromophenyl)-2-phenylethene-1,1-diyl) bis((decyloxy)benzene) (2).** Potassium carbonate (1.6 g, 11.3 mmol) and compound **1** (1 g, 2.25 mmol) were added into a 100 ml two-necked round-bottom flask. The flask was vacuumed and filled with nitrogen three times. After purging with nitrogen, 1-bromodecane (2 g, 9.04 mmol) and DMF (32 ml) were added to the flask. The reaction was stirred overnight under nitrogen conditions at 70 °C. After the mixture cooled to room temperature, the system was extracted with dichloromethane (DCM) and washed with distilled water three times and dried with anhydrous magnesium sulfate. The crude product was purified by silica column chromatography using hexane and ethyl acetate (10:1 v/v) as the elusion solvent to give 2 as a pale-yellow viscous oil (1.4g, 1.93mmol, yield:86%). $^{1}$H NMR (400 MHz, Chloroform-$d$) δ 7.20 (m, 2H), 7.04 (m, 6H), 6.89 (dtd, 6H), 6.62 (m, 4H), 3.86 (dt, 4H), 1.72 (ddd, 4H), 1.42 (q, 4H), 1.28 (m, 30H), 0.88 (t, 8H). $^{13}$C NMR (101 MHz, Chloroform-$d$) δ 157.89, 157.80, 157.65, 144.39, 143.90, 143.41, 140.97, 137.66, 136.16, 135.82, 135.73, 133.06, 132.53 (d, $J$ = 3.1 Hz), 131.37 (d, $J$ = 2.5 Hz), 130.83, 127.79, 127.64, 126.23, 125.98, 119.95, 113.71, 113.52 (d, $J$ = 3.0 Hz), 67.84 (d, $J$ = 5.1 Hz), 31.91, 29.72, 29.58 (d, $J$ = 1.7 Hz), 29.44 (d, $J$ = 2.0 Hz), 29.32 (d, $J$ = 2.0 Hz), 26.08 (d, $J$ = 1.6 Hz), 22.69, 14.13.

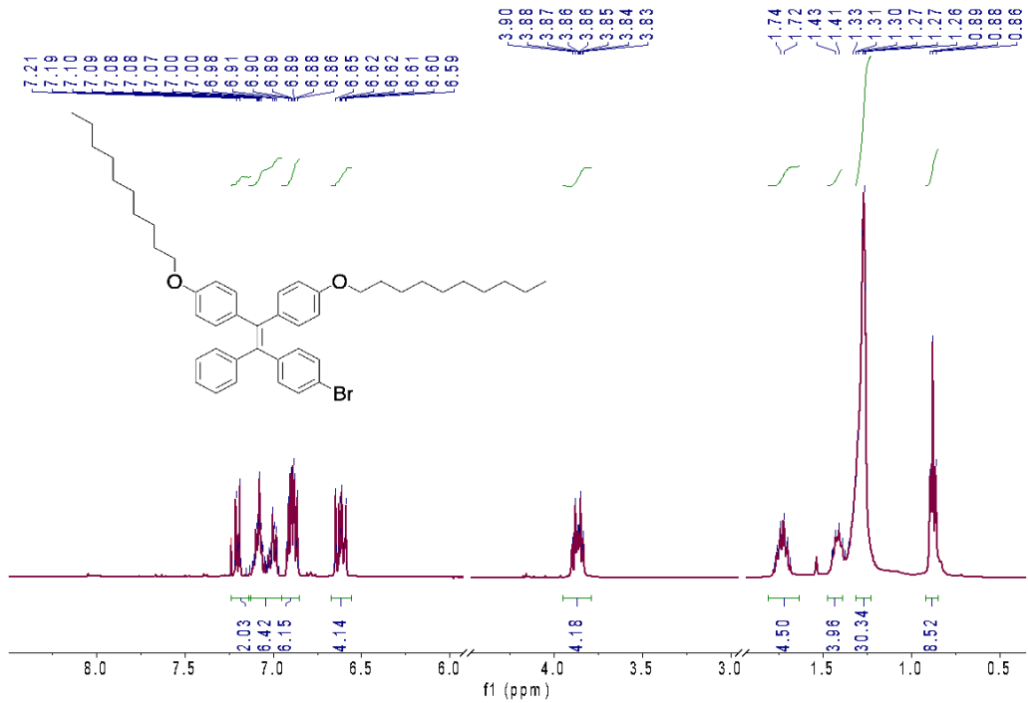

**Figure S2.** $^1$H NMR of compound 2 in Chloroform-*d*

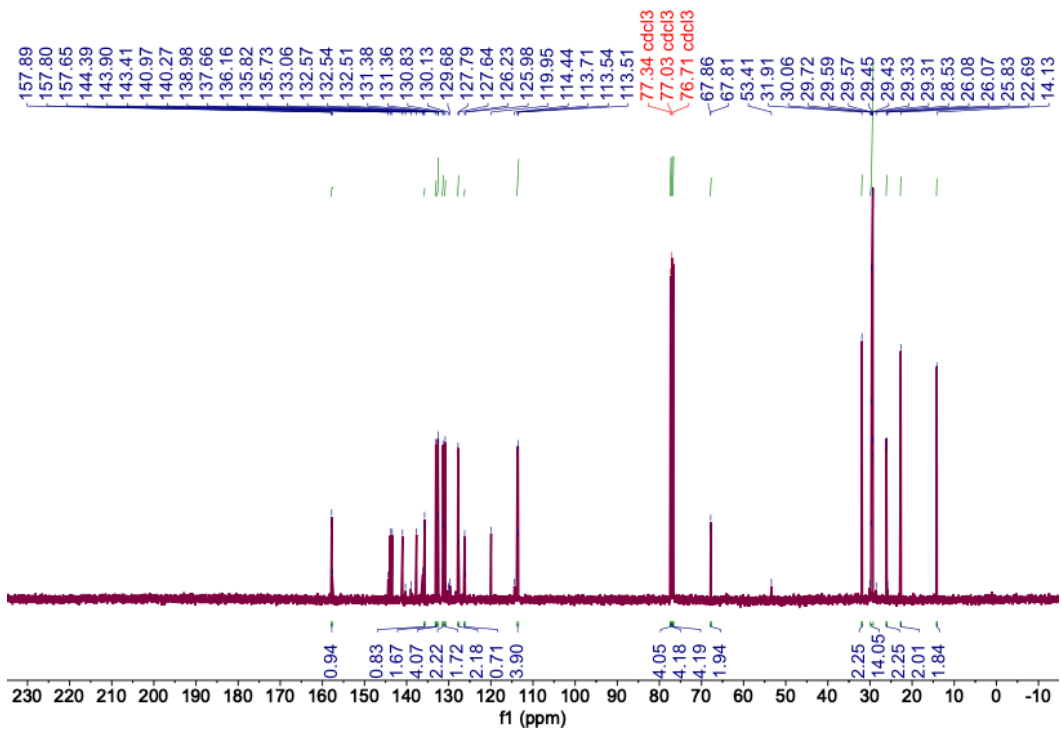

**Figure S3.** $^{13}$C NMR of compound 2 in Chloroform-*d*

**Synthesis of Methyl (Z)-4'-(2-(4-(decyloxy) phenyl)-2-(4-(nonyloxy)phenyl)-1-phenylvinyl)-[1,1'-biphenyl]-4-carboxylate (3).** Compound **2** (1.4 g, 1.93 mmol), 4-(Methoxycarbonyl) benzeneboronic acid (0.42g, 2.3 mmol), Pd (PPh$_3$)$_4$ (220 mg, 0.21 mmol) and K$_2$CO$_3$ (1.05 g, 7.6 mmol) were added into a 100 ml round-bottom flask. The flask was fitted on the Schlenk line, vacuumed, and purged with nitrogen three times. A mixture of dioxane and water (40 ml:10 ml) was bubbled with nitrogen for 30 min and then transferred to the flask using a canula. The mixture was then allowed to react for 24 hours at 100 °C. After cooling to room temperature, the mixture was poured into water and the pH was adjusted to about 5. Then, the mixture solution was extracted with DCM and washed with water three times. The organic phase was then removed and was purified by silica column chromatography using hexane and ethyl acetate (10:1 v/v) as the elusion solvent to give compound 3 as a white-yellowish solid (0.5 g, 0.64 mmol, yield :33%). $^1$H NMR (400 MHz, Chloroform-*d*) δ 8.06 (d, 2H), 7.62 (d, 2H), 7.38 (d, 2H), 7.08 (m, 7H), 6.94 (dd, 4H), 6.63 (t, 4H), 3.92 (d, 3H), 3.87 (t, 4H), 1.73 (p, 2H), 1.41 (m, 4H), 1.27 (m, 27H), 0.88 (td, 6H). $^{13}$C NMR (101 MHz, Chloroform-*d*) δ 167.03, 157.81, 157.73, 145.18, 144.54, 144.22, 140.86, 138.27, 137.08, 136.02, 132.59, 131.97, 131.44, 130.01, 129.56, 128.58, 128.34, 127.74, 126.65, 126.37, 126.14, 113.64, 113.51, 77.32, 77.00, 76.69, 67.84, 52.07, 31.88, 29.55, 29.41, 29.30, 26.05, 22.67, 14.10.

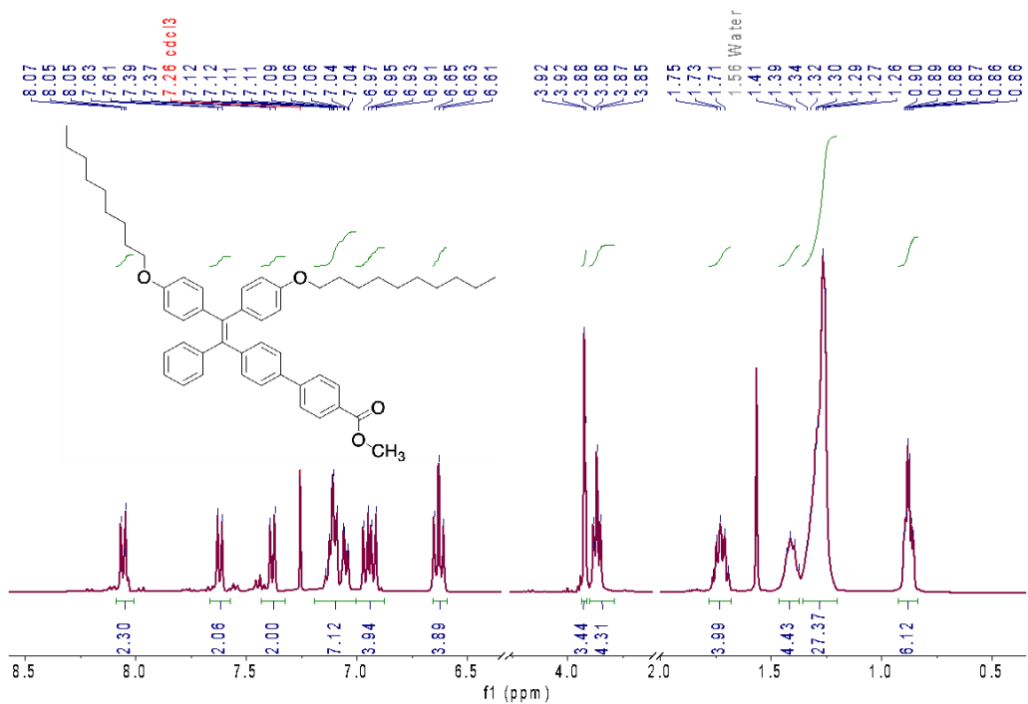

**Figure S4.** $^1$H NMR of compound 3 in Chloroform-*d*

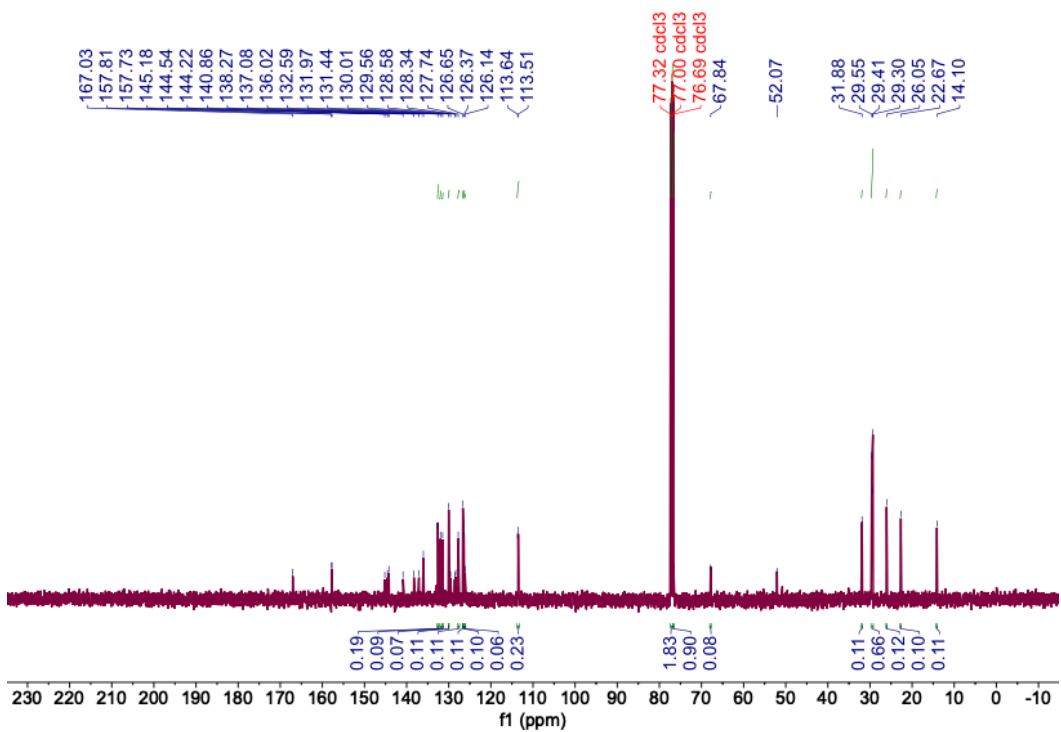

**Figure S5.** $^{13}$C NMR of compound 3 in Chloroform-*d*

**Synthesis of (*Z*)-4'-(2-(4-(decyloxy) phenyl)-2-(4-(nonyloxy) phenyl)-1-phenylvinyl)-[1,1'-biphenyl]-4-carboxylic acid (4).** 15 ml of 2M solution of sodium hydroxide in distilled water was added into a mixture of THF and MeOH (1:1 v/v) (15ml THF and 15ml MeOH). Compound **3** (0.5 gr, 0.64 mmol) was added to the mixture and allowed to reflux overnight at 75 °C. After cooling, the organic solvent was removed, and the aqueous phase was acidified with 6 M hydrochloric acid to precipitate. The precipitate was washed with water several times and dried under vacuum to give compound 4 as a dark yellow solid (0.37 g, 0.48 mmol, yield:75 %). $^1$H NMR (400 MHz, Chloroform-*d*) δ 8.13 (d, 2H), 7.65 (d, 2H), 7.39 (d, 2H), 7.09 (m, 7H), 6.95 (dd, 4H), 6.63 (t, 4H), 3.87 (t, 4H), 1.73 (p, 4H), 1.41 (t, 4H), 1.28 (dd, 25H), 0.88 (t, 6H). $^{13}$C NMR (101 MHz, Chloroform-*d*) δ 171.40, 157.69, 157.60, 145.83, 144.56, 144.06, 140.78, 138.10, 136.83, 135.87, 132.45, 131.87, 131.31, 130.51, 127.61, 126.61, 126.29, 126.01, 113.51, 113.37, 77.18, 76.86, 76.54, 67.68, 31.74, 29.41, 29.27, 29.16, 25.91, 22.53, 13.96.

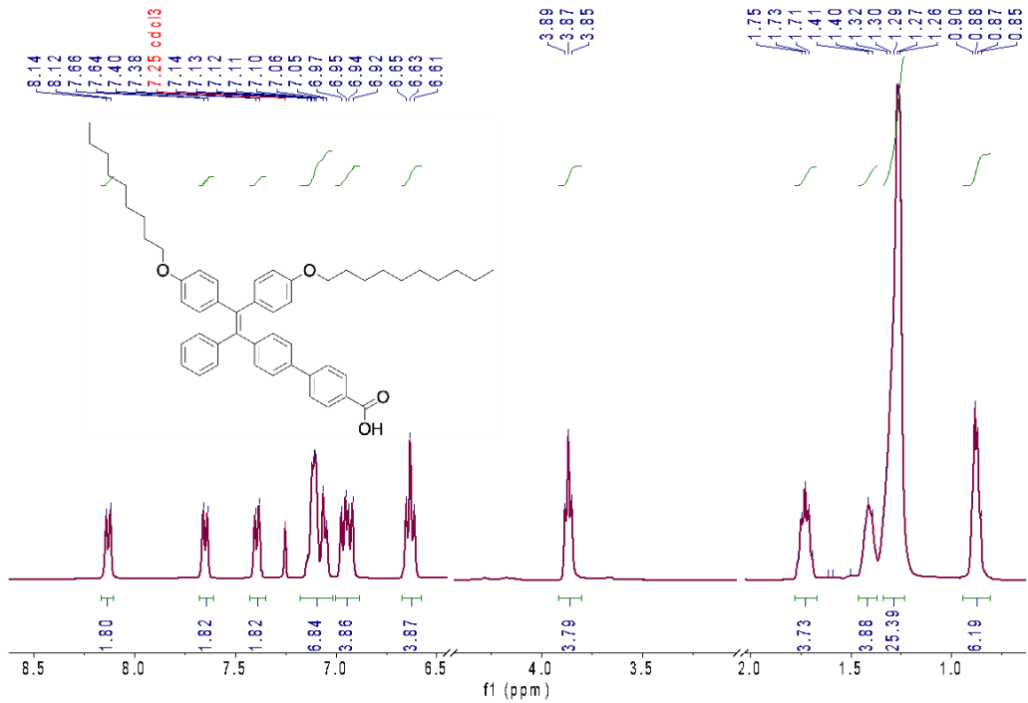

**Figure S6.** $^1$H NMR of compound 4 in Chloroform-*d*

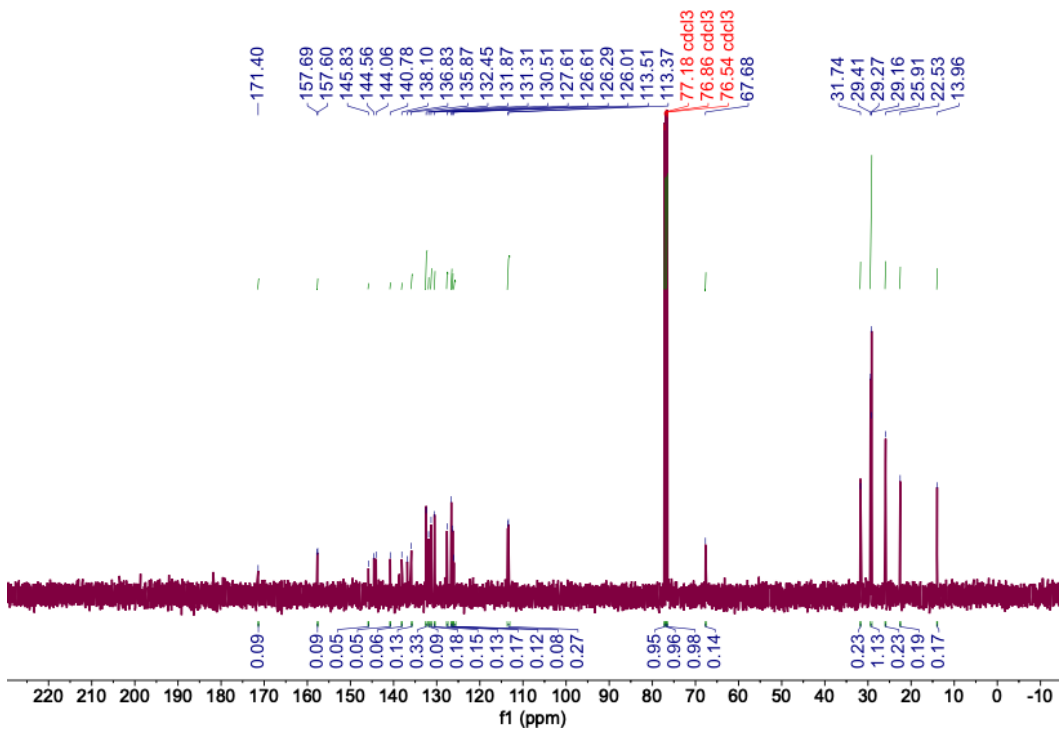

**Figure S7.** $^{13}$C NMR of compound 4 in Chloroform-*d*

**Synthesis of TPE-NHS.** Compound **4** (370mg, 0.48 mmol), pyridine (0.18 ml), and N,N'-disuccinimidyl carbonate (132.9 mg, 0.52 mmol) were dissolved in Acetonitrile (20 ml). The reaction was refluxed over night at 82 °C. After cooling, the organic solvent was removed, and the crude product was purified by a silica gel column using hexane and ethyl acetate (10:1 v/v) as eluent. After drying under vacuum, TPE-NHS ester (80 mg, 0.1 mmol, yield: 20%) was obtained as a yellow viscous oil. $^1$H NMR (400 MHz, Chloroform-*d*) δ 8.15 (d, 2H), 7.69 (d, 2H), 7.40 (d, 2H), 7.09 (dd, 7H), 6.95 (dd, 4H), 6.64 (m, 4H), 3.87 (d, 4H), 2.92 (s, 4H), 1.73 (m, 4H), 1.41 (d, 4H), 1.26 (m, 29H), 0.88 (d, 7H). $^{13}$C NMR (101 MHz, Chloroform-*d*) δ 169.32, 157.85, 136.44, 132.61, 132.08, 131.44, 131.04, 127.77, 127.04, 126.49, 113.63, 113.47, 77.32, 77.00, 76.69, 67.80, 40.96, 31.88, 29.69, 29.54, 29.41, 29.30, 26.04, 25.68, 22.67, 14.12. (MALDI-TOF) m/z: calculated, 862.16 [M]; found, 862.51 [M]$^+$.

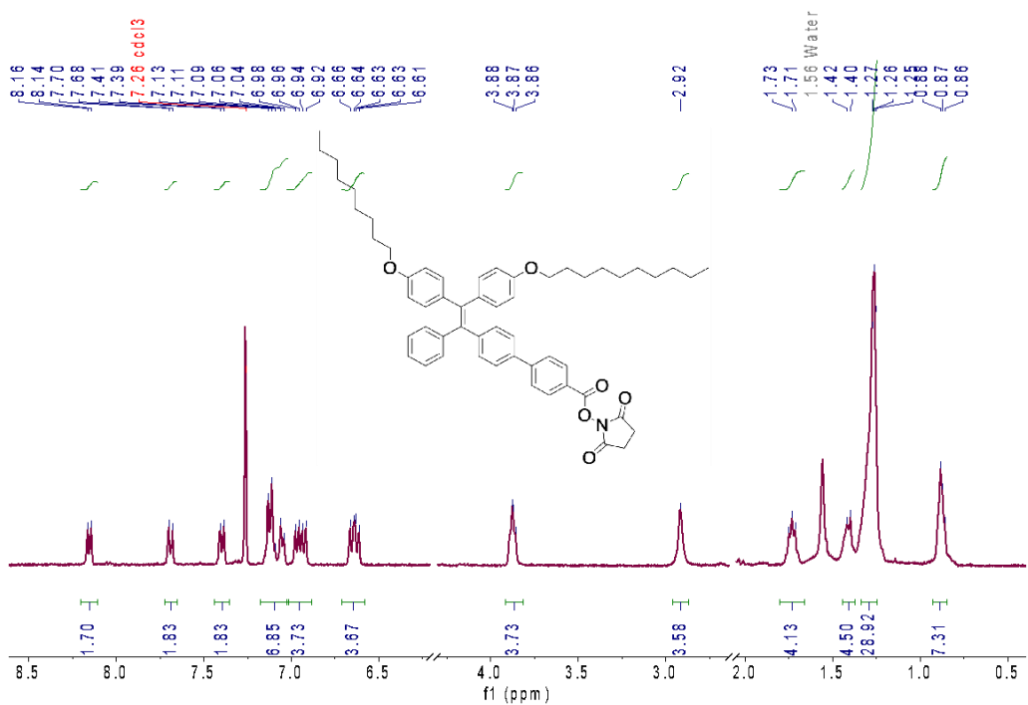

**Figure S8.** $^1$H NMR of TPE-NHS in Chloroform-*d*

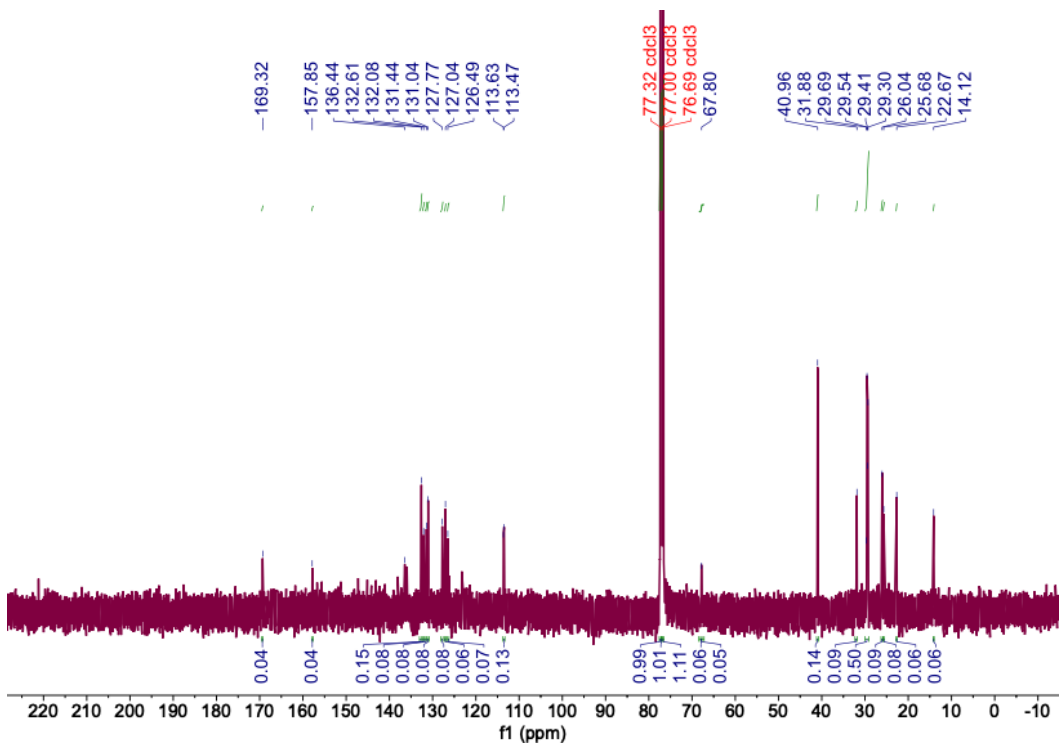

**Figure S9.** $^{13}$C NMR of TPE-NHS in Chloroform-*d*

## 3 Optical properties of TPE-NHS
### 3.1. Optical absorption and emission

A stock solution of TPE-NHS in DMSO with a concentration of 10 mM was prepared and stored at 4 °C for making all the solutions that were used in the optical characterization studies. The absorption and emission spectrum of 100 µM TPE-NHS was measured in DMSO, 99% distilled water, and 99% DMEM cell media. The absorption data was collected on Beckman Coulter DU 730 UV/Vis Spectrophotometer. The emission data was collected using HORIBA Fluoromax-4 spectrofluorometer by exciting the sample at 365 nm with excitation and emission slit width of 5nm. The background was removed using the relevant solvent. All spectra were normalized to the maximum and plotted in **Figure S10.** The change in absorption and emission wavelengths in the three different solvents can be attributed to solvatochromism.

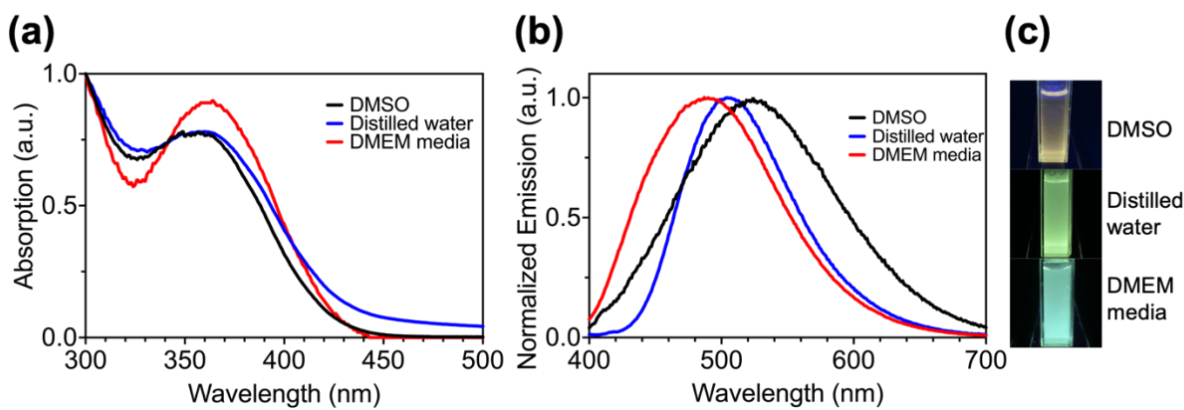

**Figure S10**. **(a)** Absorption and **(b)** fluorescent emission spectroscopy of TPE-NHS in different solvents (DMSO, distilled Water, and DMEM (cell media)) with excitation at 365 nm. **(c)** 100 µM solution of TPE-NHS in DMSO, distilled water, and DMEM media excited with a 365 nm UV lamp. All the spectra are normalized to maximum.

### 3.2. Aggregation induced emission

Two approaches were taken to initiate compound aggregation in a controlled manner. The first method changed the concentration of TPE-NHS in 99 (v/v) % distilled water/DMSO solution, and the second method changed the solubility of TPE-NHS in the solvent by increasing the relative volume ratio of water:DMSO. The emission data was collected using HORIBA Fluoromax-4 spectrofluorometer by exciting the sample at 365 nm with excitation and emission slit width of 5 nm.

First, the concentration of the TPE-NHS molecule in the 99 (v/v) % solution of distilled water/ DMSO was systematically increased from 0.1 µM to 100 µM, resulting in aggregation. TPE-NHS has a N-Hydroxy succinimide side group (mildly polar) and two long nonpolar alkane arms. The presence of these two side chains gives the compound a mildly amphiphilic property. At lower TPE-NHS concentrations, the solution remains non-fluorescent; however, as the concentration increases, aggregation and possibly micelle formation takes place, and the emission of TPE-NHS turns on. Maximum photoluminescence (PL) intensity at each concentration was plotted against TPE-NHS concentrations. The trend of fluorescence changes of TPE-NHS upon aggregation in **Figure S11a** gives 10 µM as the estimated critical aggregation concentration of the compound. When the concentration is below 1 µM, TPE-NHS dissolves in 99% distilled water solution and remains nonfluorescent. Approaching and exceeding 10 µM concentrations, the TPE-NHS molecules form aggregates which restrict intramolecular motions and result in a noticeable increase in the emission of TPE-NHS.

For the second approach, solubility testing confirmed that TPE-NHS is soluble in mildly polar or non-polar solvents such as Dimethyl sulfoxide (DMSO), Hexane, and Chloroform which have a relative polarity of 0.444, 0.009, and 0.259 respectively[2]. As shown in **Figure S11b**, because of being soluble in DMSO, the TPE-NHS solution in DMSO (100 µM) exhibited weak fluorescence centered at 500 nm. As the volume ratio of the highly polar distilled water was increased, the solution demonstrated a fluorescent turn-on process due to the formation of aggregates and the restriction of intermolecular movements of the compound. Finally, at the distilled water volume fraction of almost 100%, the emission intensity of the mixture reached its maximum.

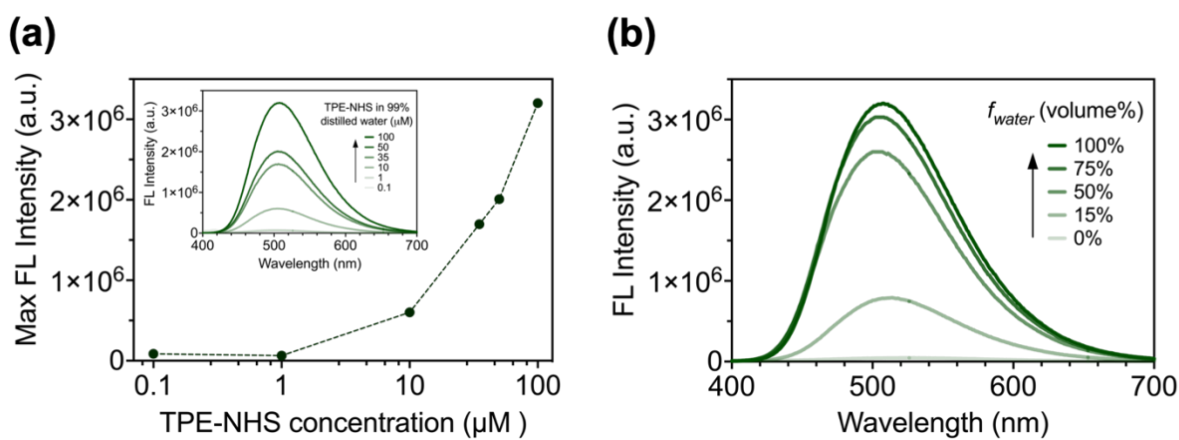

**Fig S11.** Aggregation induced emission of TPE-NHS. **(a)** Fluorescent intensity of different concentrations of TPE-NHS ester in 99% distilled water and **(b)** Fluorescent intensity of a 100 µM solution of TPE-NHS as a function of the polarity of the solvent (distilled Water: DMSO (v/v) %) (Excitation=365 nm).

### 3.3. Effect of Trastuzumab on TPE-NHS emission

The impact of Trastuzumab on the fluorescent intensity of the TPE-NHS was tested by preparing 0 µg/ml to 100 µg/ml solutions of Trastuzumab in 1x PBS and adding a constant amount of TPE-NHS in DMSO to all of them to reach the final concentration of 10 µM TPE-NHS in 99% 1x PBS. The emission spectrum of each solution was collected using HORIBA Fluoromax-4 spectrofluorometer by exciting the sample at 365 nm with excitation and emission slit width of 5nm (**Figure S12a**). To determine the maximum intensity value, the mean of the 50 largest intensity values of each spectrum was calculated. The maximum intensity of each spectrum was plotted against the Trastumzumab concentration in **Figure S12b**.

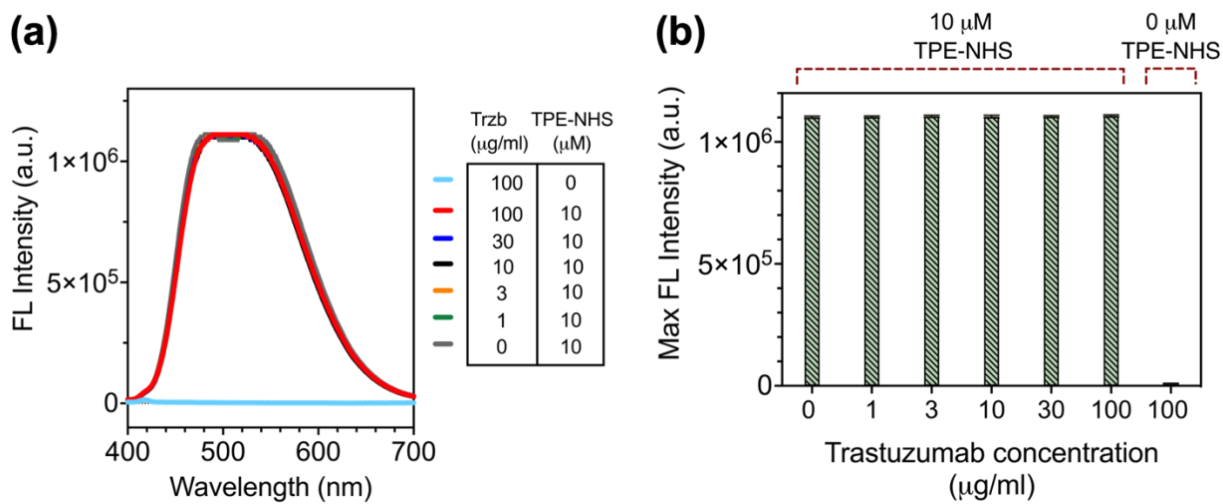

**Figure S12. (a)** Fluorescent spectrum of 10 μM TPE-NHS in 99% 1x PBS exposed to a range of Trastuzumab. (Trzb) concentrations from 0 μg/ml to 100 μg/ml and **(b)** Maximum fluorescent intensity of the fluorescent spectra of each sample plotted against its Trastuzumab concentration (Excitation=365 nm). In some cases, the error bars are not visible because they are smaller than the symbols.

## 4   Bioconjugation optimization

TPE-NHS was conjugated to HER2 antibody based on the schematic in **Figure S13**.

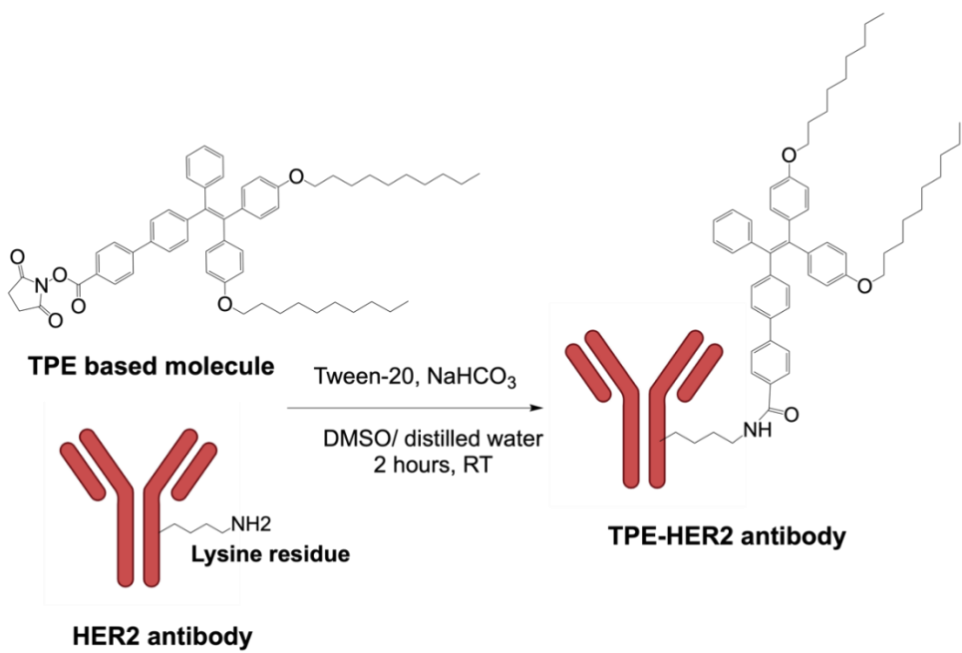

**Figure S13.** Schematic of the TPE-HER2 Ab bioconjugation reaction

To determine the optimum amount of Tween-20 for the conjugation of the HER2 antibody (Ab) to the TPE-NHS in the micelle-mediated conjugation reaction, the reaction was performed using a range of volume % of Tween-20 in the reaction solution.

The stock solutions of Tween-20 (Sigma 9005-64-5) in distilled water at various volume percentages of 0.015%, 0.03%, 0.06%, 0.12%, and 0.25%, 0.5%, and 1% were prepared by serial dilution. HER2 antibody (Santa Cruz Biotechnology, Anti-neu/ErbB2/HER2 Antibody (9G6): sc-08) was further concentrated using an ultra-centrifugal filter (Millipore Sigma Amicon ultra- 0.5 centrifugal filter devices, 10 kDa molecular weight cut off) at 14,000 xg for 40 minutes. TPE-HER2 Ab conjugation reaction was performed by mixing 35 µl of a 1.43 mg/ml solution of HER2 antibody in 1x PBS, 5 µl of 1M $NaHCO_3$ solution in distilled water, 5 µl of each Tween-20 stock solution, and 5 µl of 0.6 mg/ml (666.7 µM) solution of TPE-NHS in DMSO respectively. The reaction mixtures were vortexed at room temperature for 2 hours and then purified using gel spin column (Thermo Scientific Zeba Desalting Columns, 0.5ml, 40 MWCO). Fluorescein-HER2 Ab which was used as control was prepared by conjugation of NHS-Fluorescein (Thermo Scientific 46410) to the same HER2 antibody according to the manufacturer's protocol. The prepared TPE-HER2 Ab and Fluorescein-HER2 Ab solutions were immediately used.

All developed TPE-HER2 Ab conjugates were tested in an optimization study based on direct immunofluorescent staining of SKBR3 cells (HER2+) and MCF7 cells (HER2-). SKBR3 cells (ATCC, HTB-30) and MCF7 cells (ATCC, HTB-22) were seeded at the density of 7,000 cells per well in a 96 well glass-bottom plate (Cellvis, P96-0-N) and incubated for 3 days before running the assay to reach the approximate confluency of 70%. After 3 days, the media was removed, and the cells were fixed using 4% Paraformaldehyde (Alfa Aesar, J62478) and blocked using 2% BSA blocking buffer (Thermo scientific 37525) for one hour. After washing the fixed cells 3 times (5 minutes each), TPE-HER2 Ab conjugates were diluted to the concentration of 40 µg/ml in 0.1% BSA solution, added to the fixed SKBR3 and MCF7 cells, and left at 4°C overnight. The samples were washed with 1x PBS 2 times and imaged on Zeiss Axio observer connected to a X-cite Series 120Q light source using a TPE specific excitation and emission filter cube (excitation of G365, BS of 395, and emission BP of 535/30).

The assay determines the optimum volume % of Tween-20 based on the staining efficacy and target specificity of the developed TPE-HER2 Ab. Results in **Figure S14a** indicate that the TPE-HER2 Ab conjugates developed in presence of 0.025% Tween-20 have the maximum staining yield in SKBR3 cells. Furthermore, the relative absence of fluorescent signal in MCF7 cells stained with the same TPE-HER2 Ab indicates the target specificity of the TPE-HER2 Ab (**Figure S14b)**.

During the purification step using the gel spin column, part of the Tween-20 might pass through the column into the TPE-HER2 Ab solution. To assess whether or not the presence of Tween-20 in the staining solution affects the staining capability of TPE-HER2 Ab, fixed SKBR3 and MCF7 cells were stained with 40 µg/ml Fluorescein-HER2 Ab in 0.1% BSA solution and 40 µg/ml Fluorescein-HER2 Ab in 0.1% BSA and 0.005% Tween-20 solution and left at 4°C overnight. The samples were washed with 1x PBS 2 times and imaged on Zeiss Axio observer connected to a X-cite Series 120Q light source using a fluorescein specific excitation and emission filter cube (the excitation BP of 500/25, BS of 515, and emission BP of 535/30).

Results in **Figure S14c** demonstrate that the presence of free Tween-20 in the staining solution of Fluorescein-HER2 Ab did not cause effects like intracellular permeabilization of Fluorescein-HER2 Ab or cellular deformation in SKBR3 cells. Furthermore, based on **Figure S14d** no fluorescent signal is observed in the MCF7 cells stained in presence and absence of Tween-20 which confirms that the presence of extra Tween-20 in the solution does not cause any non-specific staining.

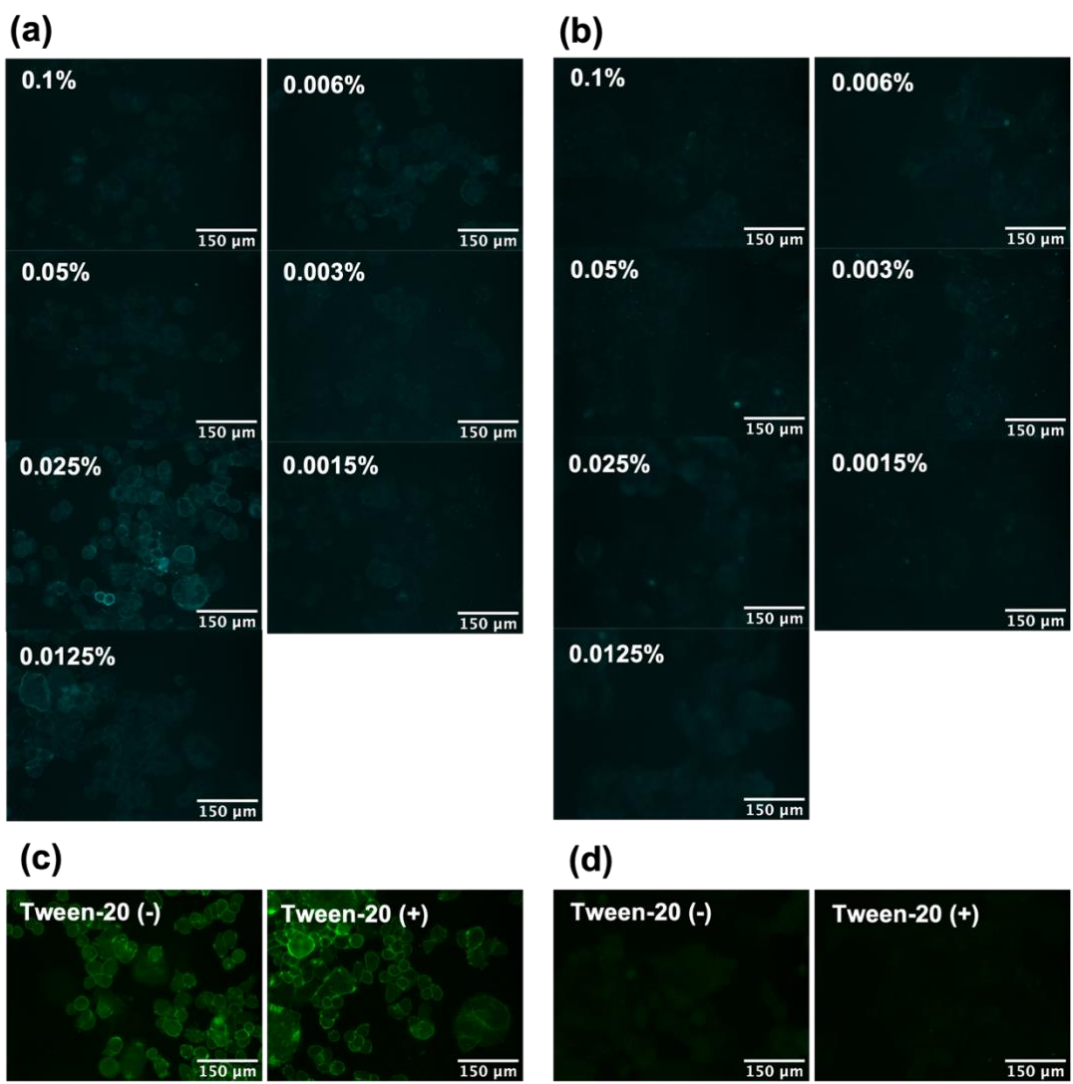

**Figure S14. (a)** SKBR3 and **(b)** MCF7 cells stained with 40 µg/ml of TPE-HER2 Ab conjugates with different amount of Tween-20 (v/v) % in the conjugation reaction solution. **(c)** SKBR3 cells and **(d)** MCF7 cells stained with 40 µg/ml of Fluorescein-HER2 Ab and 40 µg/ml of Fluorescein-HER2 Ab +0.005% Tween-20 in the staining solution.

## 5  Confirming the conjugation of TPE-HER2 Ab

Two different routes were used to confirm the conjugation reaction: MALDI mass spectroscopy and SDS-PAGE assay.

To perform the MALDI mass spectroscopy measurement, 0.7 µL of 1 mg/ml solution of HER2 Antibody and 1 mg/ml solution of TPE-HER2 Ab in PBS were spotted on a MALDI plate and allowed to dry. The salts in each sample were washed away by dispensing and removing 2 µL of distilled water several times and was left to dry. Then the dry spot was covered with 0.7 µL of 40 mg/mL 2′,6′-Dihydroxyacetophenone (DHAP) in 50% ACN, 0.1% formic acid, allowed to dry and then analysed. Samples were analysed using a Bruker Rapiflex MALDI-TOF and processed using Bruker Flex Analysis software. Results of the MALDI analysis are shown in **Figure S15**.

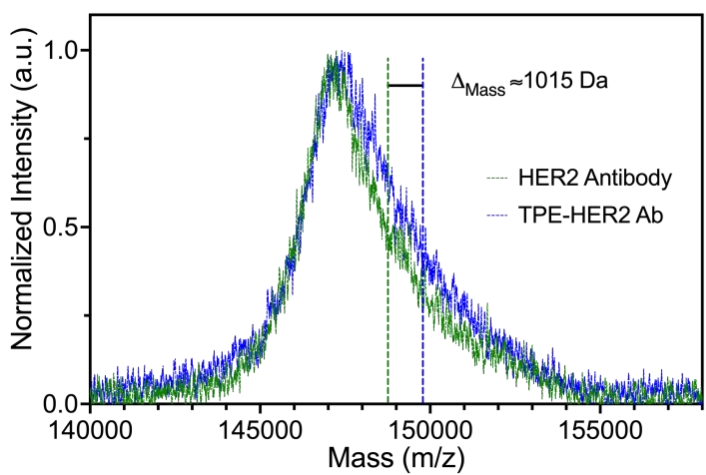

**Figure S15.** MALDI mass Spectroscopy of HER2 antibody and TPE-HER2 Ab. Mass shifted from 148,765 Da to 149,780 Da was detected by the analysis software.

To further confirm the covalent conjugation of the TPE-NHS to the HER2 antibody, an assay based on gel electrophoresis of the antibodies was designed.

The conjugated antibodies (TPE-HER2 Ab), HER2 antibody (negative control), and TPE-NHS dissolved in a 90% distilled water/DMSO solution were reduced using 1x Bolt Sample Reducing Agent (Invitrogen- B0009) and mixed with 1x Bolt LDS Sample Buffer (Invitrogen- B0007) and distilled water to get 30µl solutions of 5µg antibodies. The reduced samples were heated at 70 °C for 10 minutes and ran through the Bolt™ 4 to 12%, Bis-Tris 1.0 mm Polyacrylamide Gel (Invitrogen, NW04127BOX) using the Bolt MES SDS Running Buffer (Invitrogen-B0002) at the set voltage of 200v for 25 minutes. The gels were read using the Trans-UV 302 nm excitation and SYPRO RUBY standard emission filter (590/110 nm) on the Bio-Rad ChemiDoc imaging system.

SDS-PAGE results in **Figure S16a** indicate the presence of fluorescent signals in the heavy chain (50 kDa) and light chain (25 kDa) regions of the TPE-HER2 Ab line. Absence of the signal in the 50 kDa and 25 kDa regions of the HER2 antibody line confirms the success of the covalent conjugation of the TPE-NHS to the HER2 antibody.

A noticeable amount of the sample in TPE-HER2 Ab samples remains at the top of the SDS-PAGE gel. Most likely, this result is because of the hydrophobicity of the TPE portion of TPE-HER2 Ab which leads to the formation of fluorescent TPE-HER2 Ab aggregates. This behavior is also observed in the control TPE-NHS, supporting this hypothesis.

To confirm the attribution of the observed fluorescent peaks to light and heavy chains of HER2 antibody, the gels were transferred to a blot. The blot was blocked with a 5% blocking reagent in 1x TBS buffer for 40 minutes at room temperature and stained with the HRP linked sheep anti-mouse secondary antibody (Fisherscientific-NXA931V,1:10,000 dilution) in 0.1% TBST buffer containing 3% blocking reagent for 1 hour at room temperature. After washing the blot with 0.1% TBST 3 times (5 minutes each), the blot was developed using SuperSignal West Femto Maximum Sensitivity Substrate (Thermofisher scientific-34095) and read on the chemiluminescence channel of the Bio-Rad ChemiDoc imaging system. The presence of the chemiluminescence signal in the 50 kDa and 25 kDa regions of the TPE-HER2 Ab and HER2 Ab lines confirms that the SDS-PAGE fluorescent signals are associated with heavy and light chain portions of the antibody (**Figure S16b)**.

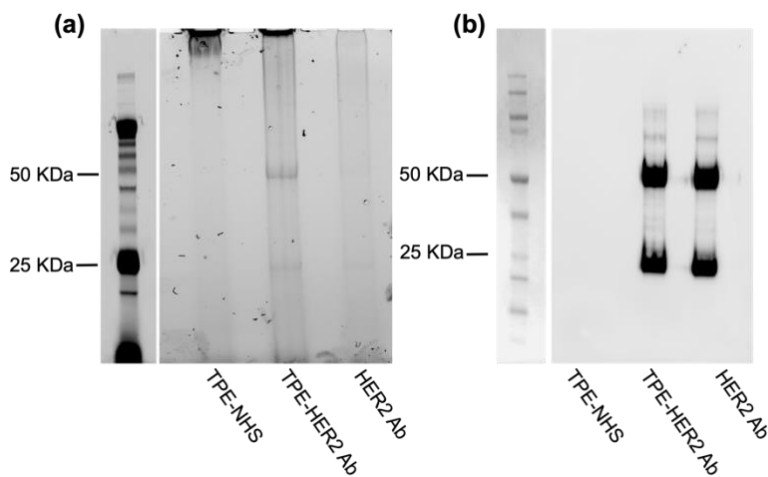

**Figure S16. (a)** SDS-PAGE gel of TPE-NHS, TPE-HER2 Ab, and HER2 Ab under reducing condition imaged in the SYPRO RUBY UV channel and displayed with inverted colors. **(b)** Western blot of the same SDS-PAGE gel including TPE-NHS, TPE-HER2 Ab, and HER2 Ab lines for confirmation of the presence of heavy and light chains of antibody in the attributed areas in the TPE-HER2 Ab and HER2 Ab lines.

## 6 Optical characterization of TPE-HER2 Ab

The absorption and emission spectrum of TPE-HER2 Ab in 1x PBS was collected on SpectraMax M2 by exciting the sample at 365 nm.

Results in **Figure S17** indicate two absorptions peaks at 280 nm and 354 nm which are attributed to the antibody portion and TPE portion of TPE-HER2 Ab respectively. The emission wavelength is centered at 518 nm.

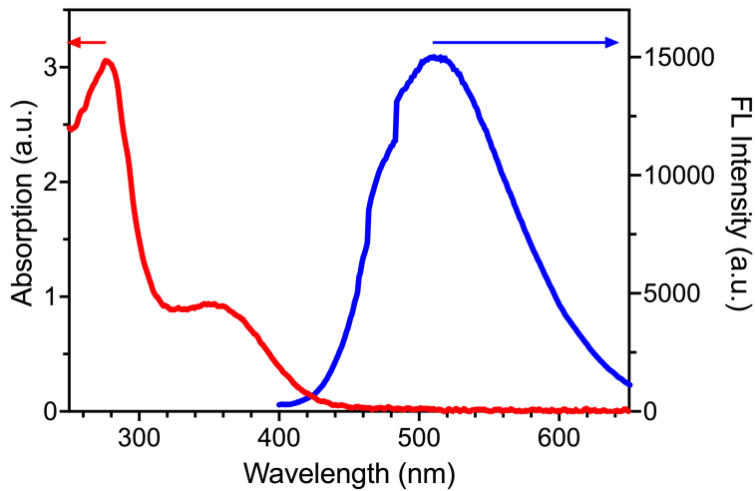

**Figure S17.** Absorption (λmax=354 nm) and emission (λmax=518 nm) spectrum of TPE-HER2 Ab

Furthermore, the AIE behavior of TPE-HER2 Ab was studied by measuring the emission spectrum of an increasing range of concentration of TPE-HER2 Ab in 1x PBS (from 0.01 mg/ml to 1.2mg/ml) as shown in **Figure S18**. To determine the maximum intensity value, the mean of the 50 largest intensity values was calculated. The maximum intensity of each spectrum was plotted against the concentration in the main text and their nonlinear correlation confirms the AIE behavior of TPE-HER2 Ab.

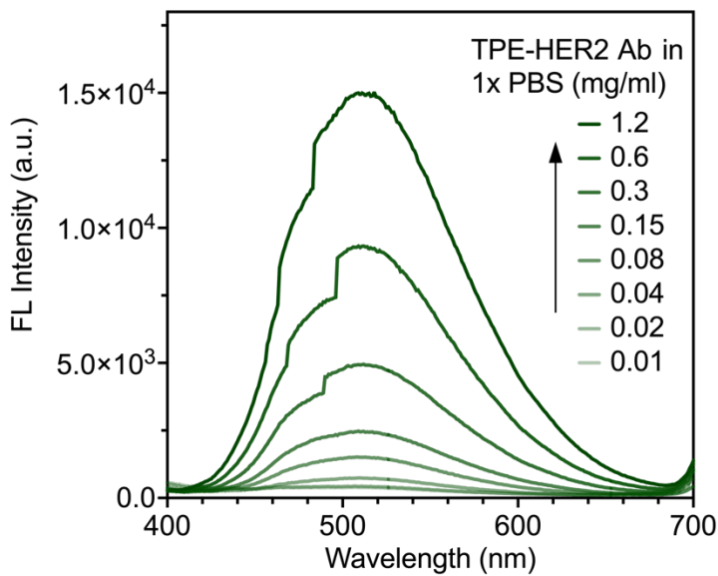

**Figure S18.** Fluorescent intensity of different concentrations of TPE-HER2 Ab in 1x PBS (Excitation=365 nm).

The discontinuity in the spectra located around 490nm in both **Figure S17** and **Figure S18** can be attributed to an issue with the grating motor in the SpectraMax M2 fluorometer. Unfortunately, because of the sample type being used (biological), an alternative fluorometer was not available. However, the absorption and emission wavelengths measured using the SpectraMax M2 agreed with those measured using the Fluoromax-4 (**Figure S11**). Therefore, given the reproducibility of the discontinuity, it can be considered to be a constant offset. Because we are interested in the relative change in fluorescent intensity as the concentration is increased, not the absolute value, ultimately, this constant offset is eliminated from the final analysis.

## 7 Cell culture

All cells were purchased from ATCC. SKBR3 cells were cultured in McCoy's 5A Medium (Gibco, 16600-082) with 10% Fetal Bovine Serum (FBS) and 1% penicillin–streptomycin at 37 °C in a humidified incubator containing 5% $CO_2$. MCF7 cells were cultured in DMEM(1x) + GlutaMAX (Gibco,10569-010) with 10% Fetal Bovine Serum (FBS) and 1% penicillin–streptomycin at 37 °C in a humidified incubator containing 5% $CO_2$. The culture medium was changed every two days and the cells were collected by treating with 0.25% (w/v) trypsin–0.53 mM EDTA solution after they reached confluence.

### 7.1. Confirmation of SKBR3 and MCF7 HER2 expression level

Prior to utilization of the SKBR3 and MCF7 cell lines in the cell-based studies, the HER2 expression level of each of these two cell lines was confirmed by indirect immunofluorescent staining and Western blotting using the same HER2 antibody that was used in the conjugation reactions.

**Indirect HER2 immunofluorescent staining and imaging of SKBR3 and MCF7 cells.** SKBR3 cells (ATCC, HTB-30) or MCF7 cells (ATCC, HTB-22) were seeded at the density of 7,000 cells per well in a 96 well glass-bottom plate (Cellvis, P96-0-N) and incubated for 3 days before running the assay to reach the approximate confluency of 70%. After 3 days, the media was removed, and the cells were fixed using 4% Paraformaldehyde (Alfa Aesar, J62478) and blocked using 2% BSA blocking buffer (Thermo scientific 37525) for one hour. After washing the fixed cells 3 times (5 minutes each), 30 μg/ml of HER2 antibody (Santa Cruz Biotechnology, Anti-neu/ErbB2/HER2 Antibody (9G6): sc-08) solution in 0.1% BSA was prepared and 50 μl of it was added to each well and incubated at 4 °C overnight. Then, wells were washed with 100 μL of 1x PBS (Phosphate Buffered Saline) for 3 times (5 minutes). The Alexa Fluor 488 rabbit anti-mouse secondary antibody (Invitrogen A11054, 1:500) solution in 0.1% BSA was prepared and 50 μl of it was added to each well and incubated for 30 minutes at room temperature. After washing the wells with 100μL of 1x PBS for 3 times (5minutes), they were imaged on FV-3000 Olympus laser scanning microscope with the excitation wavelength of 488 nm and emission detection range of 500-540 nm.

**Western blot assay of SKBR3 and MCF7 cell lysates.** Cells were seeded on a 10 cm plate until they reached 80% confluency. Then, media was removed, and cells were washed with 1x PBS three times. Protease/phosphatase inhibitor and EDTA (Thermo Scientific #78446) were added to the RIPA lysis buffer (Sigma Aldrich R0278) according to the product protocol and 1ml of prepared lysis buffer was added to each well for five minutes on ice. Cells were collected using cell scraper and moved to a 1.5 ml tube. A syringe (27-28G) was used to lyse even more of the cells by mixing 10 times. The samples were incubated on ice for 30 minutes and centrifuged at 14,000 rpm for 15 min. Supernatant was removed, and the concentration of the extracted proteins were determined using Pierce BCA assay kit (Thermo Scientific 23225 and 23227). SKBR3 and MCF7 lysed cell proteins were stored at -20°C prior to western blotting.

Each lysed cell protein extract sample was reduced using 1x Bolt Sample Reducing Agent (Invitrogen- B0009) and mixed with 1x Bolt LDS Sample Buffer (Invitrogen- B0007) and distilled water to get 30 µl solutions of 20 µg of total proteins per sample. The reduced samples were heated up at 70 °C for 10 minutes and ran through the BoltTM 4 to 12%, Bis-Tris 1.0 mm Polyacrylamide Gel (Invitrogen, NW04127BOX) using the Bolt MES SDS Running Buffer (Invitrogen-B0002) at the set voltage of 200 V for 20 minutes.

The gels were transferred to a blot. The blot was blocked with a 5% blocking reagent in 1x TBS (Tris Buffered Saline) buffer for 40 minutes at room temperature. Then the blot was incubated with separate primary antibody solutions of HER2 antibody (1:1000) and Tubulin antibody (1:5000) in 0.1% TBST buffer (1x Tris Buffered Saline with 0.1% Tween-20) containing 3% blocking reagent at 4 °C overnight. After removing the primary antibody solutions, the blots were washed with 0.1% TBST buffer 3 times (5 minutes each) and stained with the HRP linked sheep anti-mouse secondary antibody (Fisherscientific-NXA931V,1:10,000) in 0.1% TBST buffer containing 3% blocking reagent for 1 hour at room temperature. After washing the blot with 0.1% TBST 3 times (5 minutes each), the blot was developed using SuperSignal West Femto Maximum Sensitivity Substrate (Thermofisher scientific-34095) and read on the chemiluminescence channel of the Bio-Rad ChemiDoc imaging system.

**Figure S19a, b** shows strong fluorescent signal in HER2 overexpressing SKBR3 cells, and **Figure S19c, d** shows low fluorescent signal in MCF7 cells, which is due to the low HER2 expression level of this cell line. The observation of the western blot protein line associated with HER2 only in SKBR3 cell samples **(Figure S19e)** further confirms this finding and validates the selection of this pair of cell lines. Thus, these measurements simultaneously confirmed the cell lines and the HER2 antibody response.

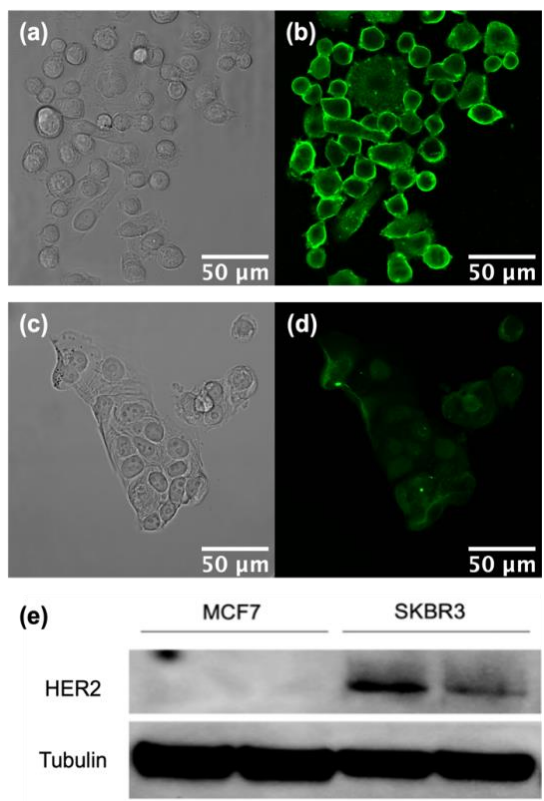

**Fig S19.** Brightfield and fluorescent channels of **(a, b)** HER2 overexpressing SKBR3 and **(c, d)** HER2 negative MCF7 cells post HER2 immunofluorescent staining and **(e)** Western blot of the HER2 overexpressing SKBR3 and HER2 negative MCF7 cell lysates in duplicates.

## 8 Cytotoxicity analysis of TPE-NHS

To study the compatibility of the system with live cells, the cytotoxicity of the TPE-NHS compound on MCF7 and SKBR3 cell lines was studied using the Cell Titer-Glo (CTG) Luminescent Cell Viability Assay (Promega). This assay determines the number of viable cells by quantitation of the ATP present, which signals the presence of metabolically active cells.

Approximately 20,000 MCF7 and SKBR3 cells were seeded in a 96 well plate and incubated for 24 hours. After 24 hours, both cell lines were treated with different concentrations of TPE-NHS which were prepared in aliquots in DMSO and were mixed by the cell media to get to the final 1% volume of DMSO in the media. Control wells including cells only, cells with 1% DMSO, and cells with 20 μM Staurosporine as positive control were also included in the staining. The treated cells were incubated for 24 hours to give the cells enough time to be exposed to TPE-NHS. After 24 hours, 50 μl of the cell media was removed and replaced with 50 μl of the CTG reagent. The contents were mixed for 2 minutes on an orbital shaker to induce cell lysis, and the plate was incubated at room temperature for 10 minutes to stabilize the luminescent signal. The luminescence signal of each well was read on the Promega GLOMAX luminometer. The assay was repeated in triplicate.

For each cell line, triplicate data was normalized with respect to the luminescence signal from the non-treated cells, combined, and plotted in **Figure S20.** The result demonstrates that TPE-NHS does not have any noticeable toxic effect on the tested cell lines over the concentration studied.

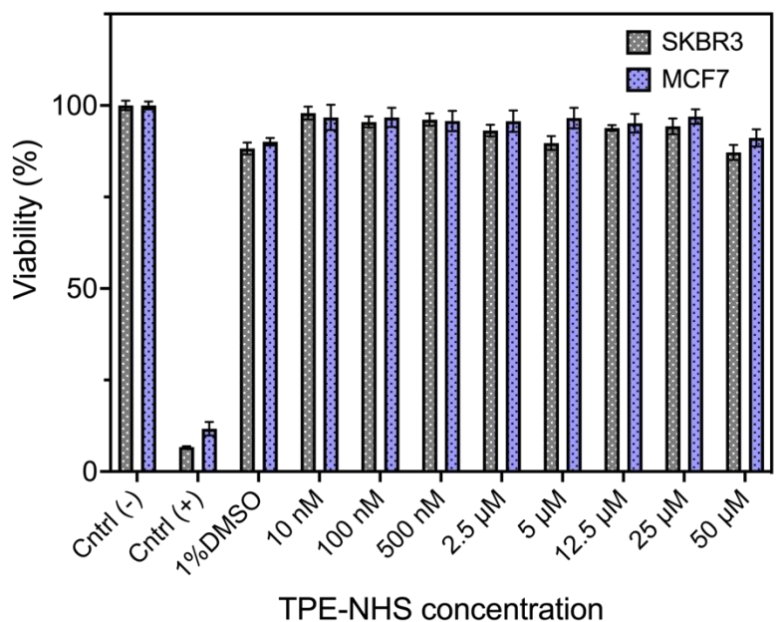

**Figure S20.** Toxicity effect of different concentrations of TPE-NHS (10 nM- 50 µM) on SKBR3 and MCF7 cells. Cntrl (-): non-treated cells or cells incubated only with cell media, and Cntrl (+): cells incubated with 20 µM Staurosporine in the media.

## 9 Signal to noise ratio

The signal to noise ratios (SNR) of the different fluorescent molecules are calculated based on the workflow described in **Figure S21** using paired brightfield and fluorescent images.

First, brightfield image of cells was used to generate a mask detecting the cell area. In this regard, the image was identified using Canny's method followed by a dilation step to close the small, disconnected boundaries (3-5 pixels) resulting from discontinuities in the edge detection. Next, the image was closed with a disk element to fill the inner cell spaces. Finally, an opening operation to remove extra dilated pixels around the cells was performed[3,4].

Subsequently, an area of the fluorescent image consisting of minimum of $3\times 10^5$ pixels including both cells and background is chosen and the defined mask is applied on selected area to filter out the cells' signal from background noise. To ensure that all fluorescent signal is omitted from the background noise, any pixel with intensity below or above the range of 'mean background signal $\pm$ 2$\times$ standard deviation of background signal' was filtered out from the background noise.

Then, SNR was calculated by dividing the mean value of the fluorescent signal and the standard deviation of the background noise of the selected area. The method is based on the approach published in the reference[5].

Each SNR value in the main text is reported in the format of 'mean$\pm$ standard deviation' of three different image replicates of the same condition.

The code is available at: https://github.com/soheilsoltani86/SNR.git

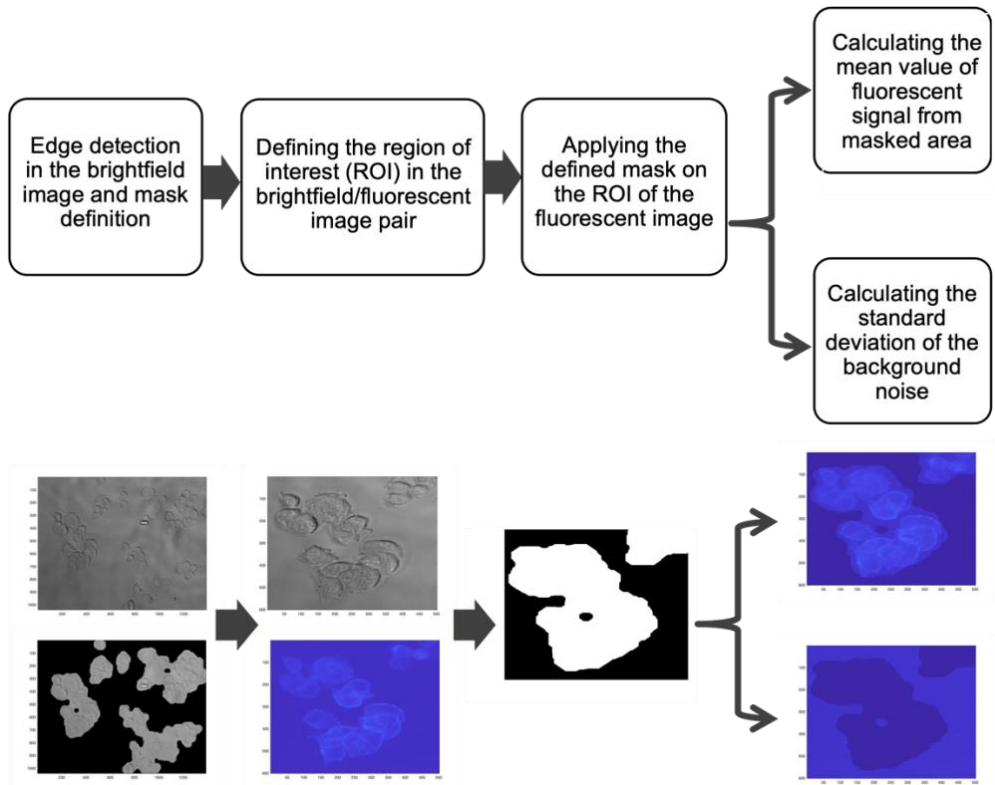

**Figure S21.** Workflow of the image analysis platform for SNR calculation from sample Images

## 10 Control imaging measurements

### 10.1. Control multi-channel fluorescent imaging

Control multi-channel fluorescent imaging measurements to confirm the optical isolation of the channels were performed.

SKBR3 cells in individual wells were stained with 20 µg/ml of Fluorescein-HER2 Ab, 20 µg/ml Texas red-HER2 Ab, and 20 µg/ml of TPE-HER2 Ab (based on the direct immunofluorescent imaging protocol in the experimental section of the main text). The Texas red-HER2 Ab stained wells were imaged in the brightfield, Fluorescein, and TPE channels of the microscope. TPE-HER2 Ab and Fluorescein-HER2 Ab stained wells were imaged in the brightfield and Texas red-HER2 Ab channels.

TPE filter cube has the excitation of G365, BS of 395, and emission BP of 535/30, the Fluorescein filter cube has the excitation BP of 500/25, BS of 515, and emission BP of 535/30, and the Texas red filter cube has the excitation BP of 550/25, BS of 570, and emission BP of 605/70.

The absence of fluorescent signal in the Fluorescein channel and TPE channel in **Figure S22b** and **Figure S22d** respectively, and the absence of fluorescent signal in the Texas-red channel in **Figure S22a** and **Figure S22c** indicates that that the fluorophores do not leak into the other channels and can be used as pairs for the colocalization study.

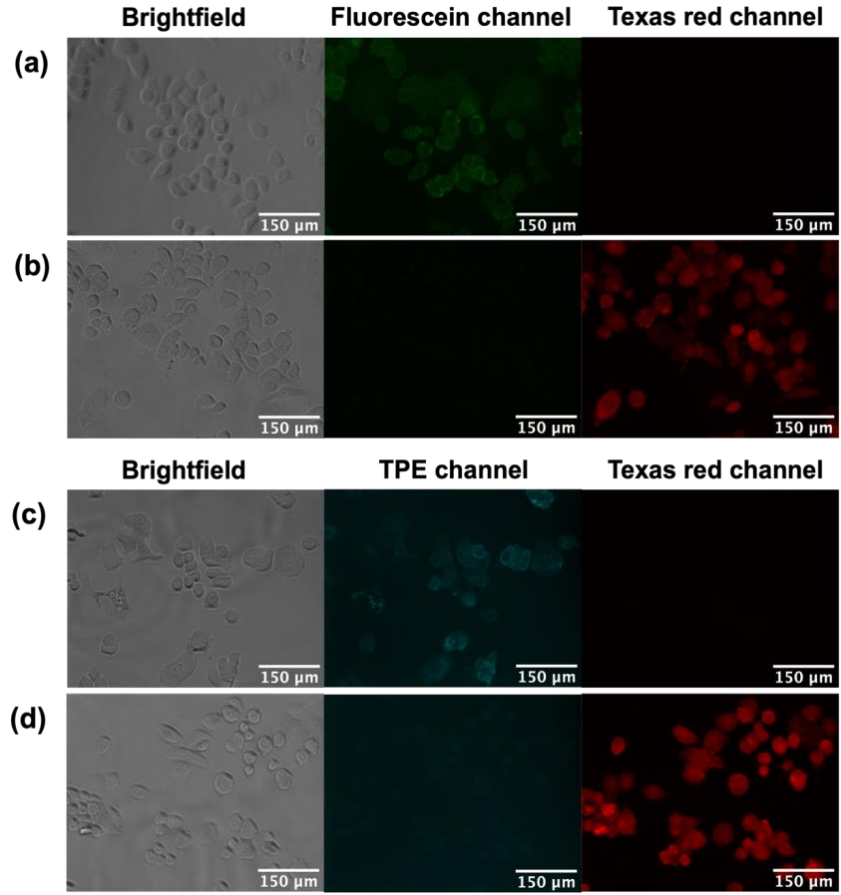

**Figure S22.** Fluorescent leaking analysis between Fluorescein channel and Texas red channel in **(a)** 20 µg/ml Fluorescein-HER2 Ab and **(b)** 20 µg/ml Texas red-HER2 Ab stained SKBR3 cells and between TPE channel and Texas red channel in **(c)** 20 µg/ml TPE-HER2 Ab and **(d)** 20 µg/ml Texas red-HER2 Ab stained SKBR3 cells.

### 10.2. Control colocalization imaging

SKBR3 were seeded at the density of 7,000 cells per well in a 96 well glass-bottom plate (Cellvis, P96-0-N) and incubated for 3 days before running the assay to reach the approximate confluency of 70%. After 3 days, the media was removed, and the cells were fixed using 4% Paraformaldehyde (Alfa Aesar, J62478) and blocked using 2% BSA blocking buffer (Thermo scientific 37525) for one hour. After washing the fixed cells 3 times, staining solution consisting of 20 µg/ml of Fluorescein-HER2 Ab and 20 µg/ml Texas red-HER2 Ab in 0.1% BSA was added to the cells. After overnight staining at 4°C, fixed SKBR3 cells were washed and imaged in the brightfield channel, Fluorescein channel (filter cube with excitation BP of 500/25, BS of 515, and emission BP of 535/30), and Texas red channel (filter cube with the excitation BP of 550/25, BS of 570, and emission BP of 605/70) on a Zeiss Axio-observer widefield fluorescent microscope.

**Figure S23a-d** shows the stained SKBR3 cells in the brightfield, Fluorescein, Texas red, and merged channel. The yellow color in the merged channel along with the linear distribution of its associated 2D histogram in **Figure 4f** in the main text represents a control standard for the TPE-HER2 Ab/Texas red-HER2 Ab colocalization assay discussed in the main text.

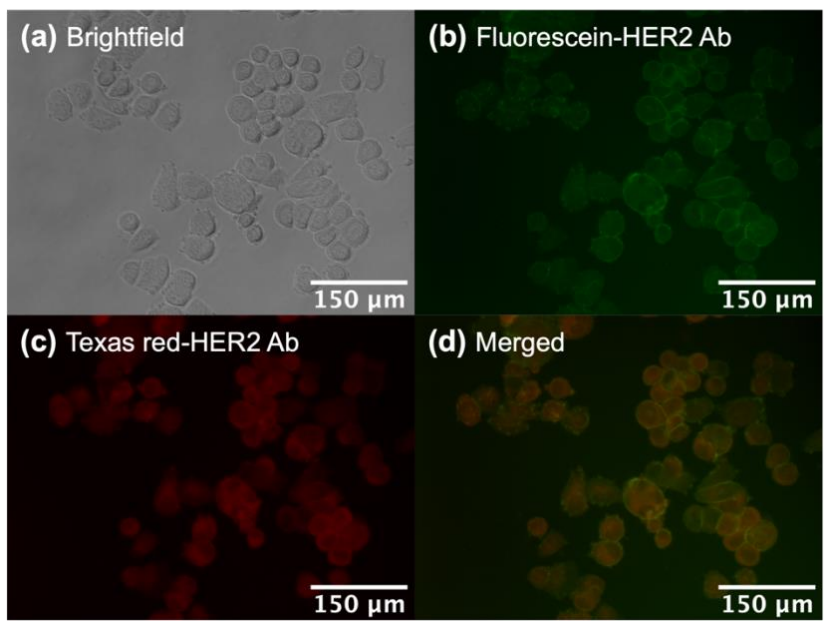

**Figure S23.** Colocalization analysis of HER2 overexpressing SKBR3 cells stained with 20 µg/ml of Fluorescein-HER2 Ab and 20 µg/ml of Texas red-HER2 Ab. Images were captured in the **(a)** Fluorescein fluorescent channel, **(b)** Texas-red fluorescent channel, **(c)** brightfield channel, and **(d)** merged image of Fluorescein and Texas red fluorescent channels.

To quantitatively assess the colocalization, the Pearson Correlation Coefficient (PCC) is calculated.

Cellular area in the brightfield channel of each colocalization data (consisting of a brightfield image and a pair of fluorescent images) were manually selected in Image J and defined as the region of interest (ROI). The ROI filters were used for masking the cellular regions in both fluorescent images of the colocalization data. Then, the masked fluorescent pairs were analyzed for calculating the PCC in Image J and plotting the 2D histogram.

Results in **Figure S24** shows the PCC values of three colocalization data replicates for TPE/Texas red colocalization and three colocalization data replicates for the control (Fluorescein/Texas red) colocalization assays. The PCC value of the TPE/Texas red colocalization assay is 0.74± 0.06 which agrees with the range of PCC for standard the control assay (0.65± 0.03).

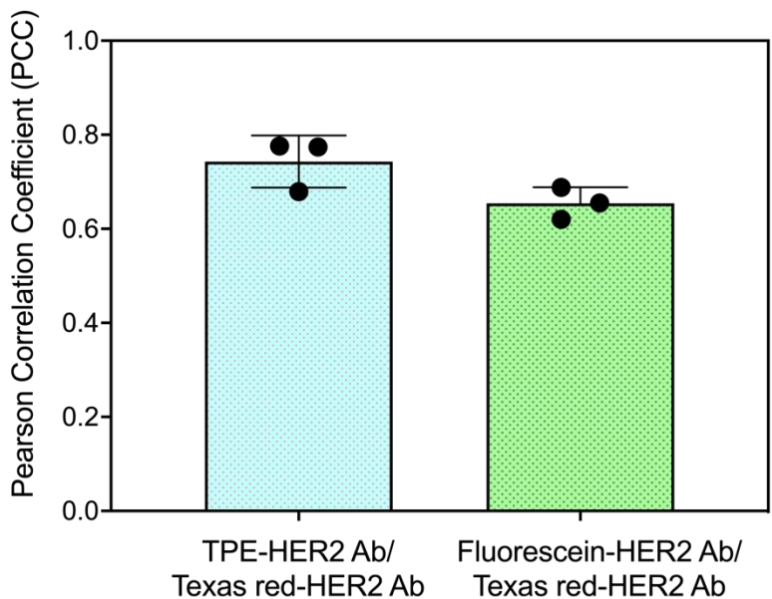

**Figure S24.** Pearson Correlation Coefficient of TPE-HER2 Ab/ Texas red-HER2 Ab stained SKBR3 cells and Fluorescein-HER2 Ab/ Texas red-HER2 Ab stained SKBR3 cells (control) in triplicates and each data point on average consists of 20 SKBR3 cells.

### 11 Image analysis platform for the AIE based assay

**Figure S25** describes the steps of the developed image analysis platform. To calculate the average intensity for treated and non-treated cells, the existing contrast in brightfield images have been used to generate a mask. Initially, the edges of the image were identified using Canny's method[3,4]. This process is followed by a dilation step to close the small, disconnected boundaries (3-5 pixels) resulting from discontinuities in the edge detection. Next, the image was closed with a disk element (Size 8 pixels) to fill the inner cell spaces. Finally, an opening operation to remove extra dilated pixels around the cells was performed. Subsequently, this mask is used in fluorescent images to filter out the cells from background. The code is available at:

https://github.com/soheilsoltani86/FLsegmentation/blob/main/masking_Yasaman_Moradi.m

The average fluorescent intensity of the masked regions of each fluorescent image is extracted from the image in a separate file and used as the comparison value between different fluorescent pictures in different conditions.

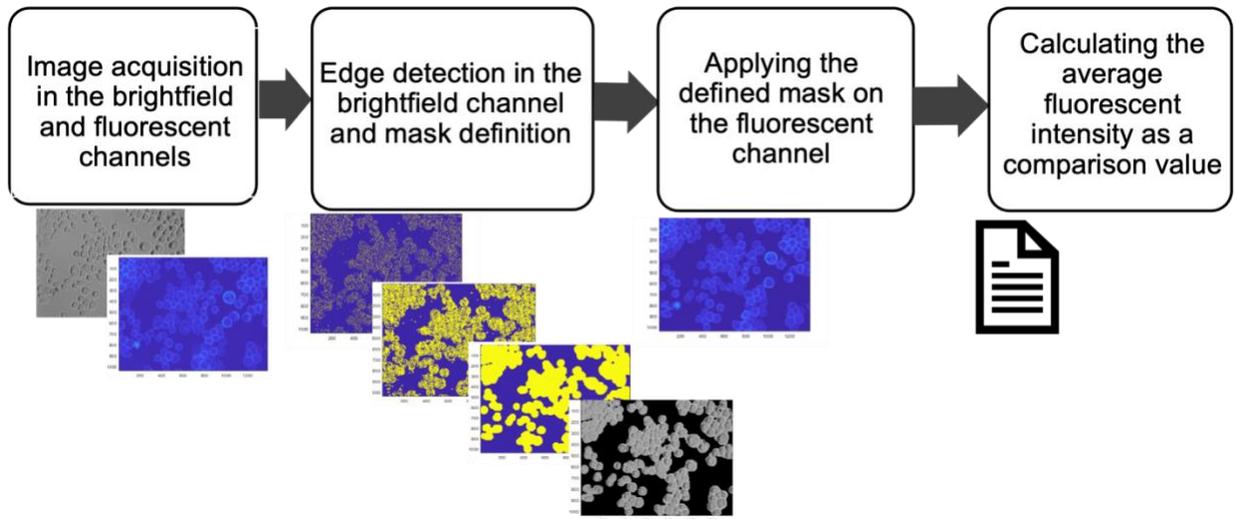

**Figure S25.** Workflow of the image analysis platform for analysis of HER2 cluster manipulation

## 12 Trastuzumab treatment

The data for all therapeutic concentrations and time measurements are shown in **Figure S26a-f**.

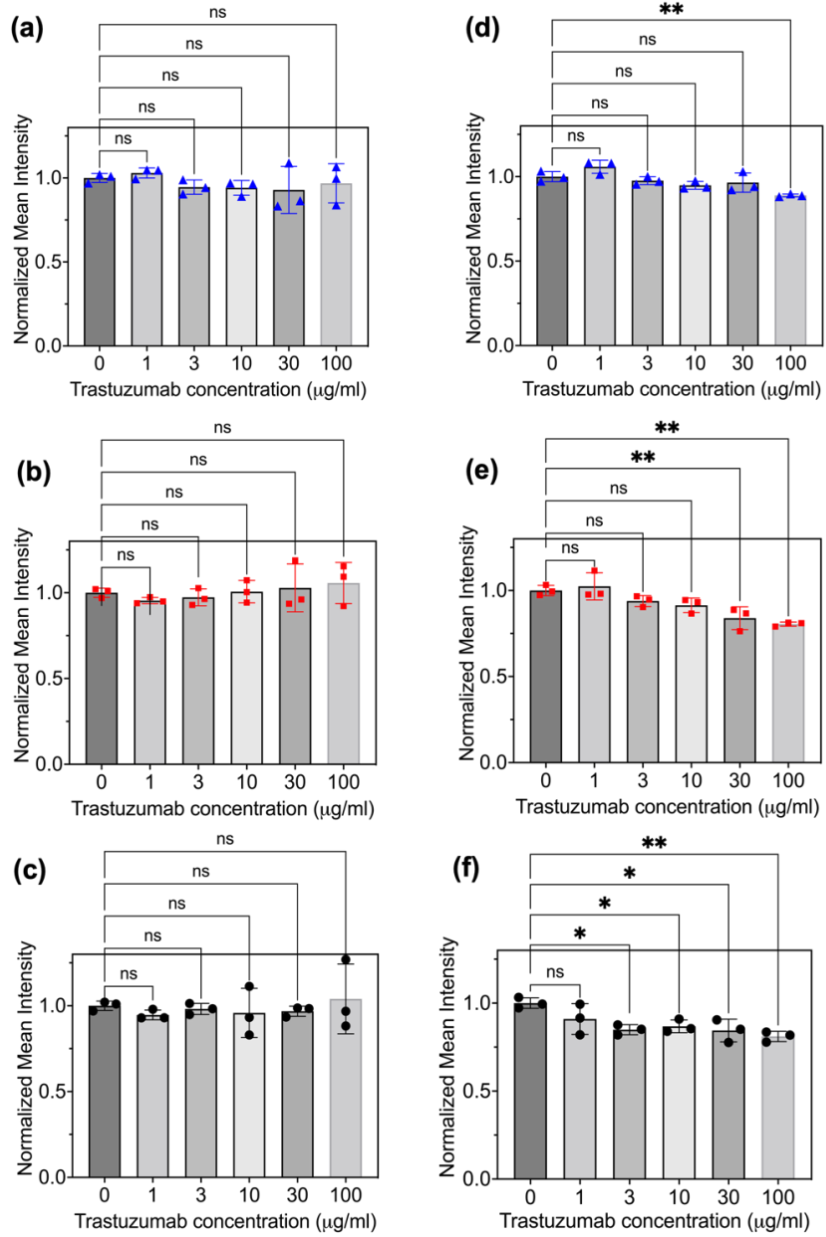

**Figure S26.** Normalized mean fluorescent intensity of SKBR3 cells stained with Fluorscenin-HER2 Ab after Trastuzumab treatment times of **(a)** 2 hours, **(b)** 8 hours, **(c)** 24 hours along with normalized mean fluorescent intensity of SKBR3 cells stained with AIE based tool after Trastuzumab treatment times of **(d)** 2 hours, **(e)** 8 hours, **(f)** 24 hours. The data is collected in triplicates and each data point on average consists of SKBR3 30 cells. (* $p < 0.05$ and ** $p < 0.01$)